\documentclass[12pt]{article}
\newsavebox{\ns}
\newsavebox{\dbrane}
\newsavebox{\dbshort}

\usepackage{epsfig}
\usepackage{amssymb}

\def\appendix{{\newpage\section*{Appendix}}\let\appendix\section%
         {\setcounter{section}{0}
         \gdef\thesection{\Alph{section}}}\section}

\newcommand\ba{\begin{eqnarray}}
\newcommand\ea{\end{eqnarray}}

\def\Dslash{\,\,{\raise.15ex\hbox{/}\mkern-12mu D}}
\def\Dbarslash{\,\,{\raise.15ex\hbox{/}\mkern-12mu {\bar D}}}
\def\delslash{\,\,{\raise.15ex\hbox{/}\mkern-9mu \partial}}
\def\delbarslash{\,\,{\raise.15ex\hbox{/}\mkern-9mu {\bar\partial}}}
\def\pslash{\,\,{\raise.15ex\hbox{/}\mkern-9mu p}}
\def\calDslash{\,\,{\raise.15ex\hbox{/}\mkern-12mu {\cal D}}}

\newcommand\re{{\rm Re}}

\newcommand{\hh}{{1\over 2}}

\renewcommand{\ll}{_}
\newcommand{\uu}{^}
\newcommand{\pp}{\partial}

\renewcommand{\exp}[1]{{\rm exp}\{#1\}}

\newcommand{\m}{\mu}

\renewcommand{\dag}{{}^\dagger{}}

\renewcommand{\m}{\mu}
\newcommand{\n}{\nu}
\newcommand{\s}{\sigma}
\renewcommand{\t}{\tau}
\newcommand{\G}{\Gamma}
\newcommand{\g}{\gamma}
\renewcommand{\a}{\alpha}

\renewcommand{\o}{\omega}
\newcommand{\e}{\epsilon}
\renewcommand{\O}{\Omega}

\newcommand{\sqd}{^2}
\newcommand{\zb}{{\bar{z}}}

\renewcommand{\hh}{{1\over 2}}

\newcommand{\eee}[1]{\ba{#1}\ea}
\renewcommand{\th}{\theta}
\renewcommand{\t}{\tau}

\renewcommand{\b}{\beta}

\newcommand{\llsk}{\hskip .5in}

\newcommand{\st}{{}^*}

\newcommand{\rmm}{r_{-}}

\newcommand{\pr}{^\prime {}}

\newcommand{\apr}{{\alpha^\prime} {}}

\newcommand{\IZ}{\relax\ifmmode\mathchoice
{\hbox{\cmss Z\kern-.4em Z}}{\hbox{\cmss Z\kern-.4em Z}}
{\lower.9pt\hbox{\cmsss Z\kern-.4em Z}} {\lower1.2pt\hbox{\cmsss
Z\kern-.4em Z}}\else{\cmss Z\kern-.4em Z}\fi} \font\cmss=cmss10
\font\cmsss=cmss10 at 7pt
\newcommand{\inbar}{\,\vrule height1.5ex width.4pt depth0pt}
\newcommand{\IC}{{\relax\hbox{$\inbar\kern-.3em{\rm C}$}}}
\newcommand{\IQ}{{\relax\hbox{$\inbar\kern-.3em{\rm Q}$}}}
\newcommand{\IP}{\relax{\rm I\kern-.18em P}}

\renewcommand{\re}{{\rm Re}}

\newcommand{\ed}{\dot{e}}

\renewcommand{\l}{\lambda}

\renewcommand{\o}{\omega}

\renewcommand{\pr}{{}^\prime{}}

\newcommand{\pst}{\tilde{\psi}}

\newcommand{\ald}{{\dot{\alpha}}}
\newcommand{\bed}{{\dot{\beta}}}
\newcommand{\Pst}{\tilde{\Psi}}

\newcommand{\pd}{{\dot{p}}}

\newcommand{\IR}{\relax{\rm I\kern-.18em R}}
\def\blfootnote{\xdef\@thefnmark{}\@footnotetext}

\renewcommand{\ss}{\nn\\}

\newcommand{\bm}{\begin{matrix}}
\renewcommand{\em}{\end{matrix}}

\newcommand{\up}[1]{^{({#1})}{}}

\newcommand{\tr}{{\rm tr}}

\newcommand{\bbb}{\ba\begin{array}{c}}
\renewcommand{\eee}{\nonumber\end{array}\ea}
\newcommand{\een}[1]{\label{#1}\end{array}\ea}

\def\hilo{{}_{{}_{{}_{{}_{{}_{}}}}} {}^{{}^{{}^{}}}}

\newcommand{\heading}[1]{\begin{center}\it {#1} \rm \end{center}}

\def\lrdd{\left ( ~}
\def\rrdd{\hilo \right )}
\def\lsqq{\left [ ~}
\def\rsqq{\hilo \right ]}

\newcommand{\kket}[1]{\left | {#1} \right \rangle }
\newcommand{\tht}{{\tilde{\theta}}}
\def\pht{{\tilde{\phi}}}
\def\bi{\begin{itemize}}
\def\ei{\end{itemize}}

\def\ed{\end{document}}

\def\tth{\tilde{\theta}}
\newcommand{\ttat}{{\tt \char`\@}}
\newcommand{\bb}[1]{{\bf {#1}}}
\def\aad{{\dot{a}}}

\def\nsm{{\rm NS}_-}
\def\nsp{{\rm NS}_+}
\def\rmm{{\rm R}_-}
\def\rpp{{\rm R}_+}
\def\rnu{{\rm \#}}
\def\ss{{\Sigma}}
\def\sst{{\tilde{\Sigma}}}
\def\aald{{\dot{A}}}
\def\pd{{\dot{p}}}
\def\mfls{(-1)^{F_{L_S}}}
\def\ttb{{\widetilde{\rm IIB}}}
\def\sstb{\bar{\tilde{\Sigma}}}
\newcommand{\rown}[9]{\bf {#1}{#1} & \bf {#2}{#2} & \bf {#3}{#3} &
\bf {#4}{#4} & \bf {#5}{#5} & \bf {#6}{#6} &
\bf {#7}{#7} & \bf {#8}{#8} & \bf {#9}{#9} }
\newcommand{\bd}[1]{{\bf {#1}{#1}}}
\newcommand{\rowf}[5]{\hline \bd{#1} & \bd{#2} & \bd{#3} & \bd{#4}
& \bd{#5}}
\newcommand{\rowfa}[6]{\hline \bd{#1},\bd{#2} & \bd{#3} & \bd{#4} & \bd{#5}
& \bd{#6}}
\def\btt{\begin{table}}
\def\ett{\end{table}}
\def\bta{\begin{tabular}}
\def\eta{\end{tabular}}
\def\ofm{{\rm O5}^-}
\begin{document}

\begin{titlepage}

\begin{center}

\hfill

\vskip 1.5 cm
{\large \bf New Type II String Theories}\\
\vskip 0.16cm
{\large \bf With}\\
\vskip 0.16cm
{\large \bf Sixteen Supercharges }\\
  \vskip 0.16cm

\vskip 1 cm
{Simeon Hellerman$^1$}\\
\vskip 1cm

$^1${\sl School of Natural Sciences,
Institue for Adanced Study, \\
Princeton, NJ 08540 \\ {\tt simeon@ias.edu}\\}

\end{center}

\vskip 0.5 cm
\begin{abstract}
We present two new backgrounds of type IIA string theory
preserving 16 supercharges.  One is a Wilson
line for $(-1)\uu{F_{L_S}}$ and the other is
an orbifold by a reflection of four coordinates, 
along with the action of $(-1)\uu{F_{L_S}}$, where
$F_{L_S}$ is left-moving spacetime fermion number.  The Wilson
line theory has many new phenomena, including a
self-duality of type IIA on a single circle, enhanced gauge
symmetry at the self-dual radius, and a T-duality between
uncharged and (locally) charged branes.  The orbifold
theory also presents many novel features, including
charged, stable non-BPS D1, D3, and D5-branes pinned
to the fixed locus, and 
an instability of the D0-brane near the fixed locus.

\end{abstract}

\end{titlepage}

\pagestyle{plain}
\setcounter{page}{1}
\newcounter{bean}
\baselineskip16pt
\tableofcontents

\newpage

\section{Introduction}
\label{intro}

String theories with large
amounts of supersymmetry have proven 
useful in probing general properties
of string dynamics in relatively
controlled models.  In this
paper we will present two 
solutions of type IIA string
theory\footnote{Though see \cite{sens}, \cite{oren} for
earlier studies of the type IIB version of our second model.
The branes of our second model in its
type IIA version 
also appear in the classification \cite{trunc}.  In the
language of \cite{trunc} they are 'truncated' branes
with $(r,s) = (1,0)$.  J. Walcher
has pointed out to us that both our models can be viewed as
orbifolds with discrete torsion.  We do not
emphasize this language, however, making instead
a direct analysis of the consistency conditions for
the choice of open and closed
string sectors.}
which preserve sixteen
supercharges.  Both theories are exactly solvable
at tree level and can be realized as
quotients of flat 10-dimensional space.

Both backgrounds involve the operation $\mfls$, which
acts as charge conjugation on all RR fields, and
as a minus sign on left-moving spin fields $\tilde{\ss}$.
The first background is a Wilson line for $\mfls$, and
the second is an orbifold of four coordinates, where
we include $\mfls$ in the orbifold action.

We will study these theories by considering first
the closed string theory, then using consistency conditions
to derive the allowed sets of boundary conditions in the
open string theory.  In the case
of the Wilson line we begin by using the Green-Schwarz
formalism in light-cone gauge to
get a quick idea of the spectrum,
then transition to the RNS formalism later.  Using the
GS formalism of course will force us to consider objects
which are at least onebranes in the noncompact directions.
When we switch to the RNS formalism we will not be forced
to adopt that restriction.

Our motivation
in studying these
backgrounds is that they are
simplified models which share some properties of
the kinds of nongeometric type II string theories studied
in \cite{Hellerman:2004zm}.  (For recent related
work, see \cite{related}.)  Namely they are
type II theories which have worldsheet
descriptions, are worldsheet-chirally asymmetric
and have nonzero but reduced supersymmetry relative to
ordinary type II models in the same number of dimensions.
Indeed, by
combining the two orbifolding operations
performed separately in this paper, one recovers a
limit of the $\hat{c} = 4$
asymmetric orbifold point described in
\cite{Hellerman:2004zm} in which three of the four directions
decompactify.

Both our theories are quite interesting
in their own right, particularly
in that all the branes we study are non-BPS but
all have at least some range of closed string moduli in
which they are stable at tree level.  In the orbifold
example, some of the non-BPS branes are always
stable, and can be related to stable, non-BPS
fundamental strings by a duality chain.
The backgrounds we
describe are fully geometric, and none of the
nongeometric ideas of \cite{Hellerman:2004zm} are necessary
to understand our results.

\section{A Wilson line for $(-1)\uu{F_{L_S}}$.}

Given any exact symmetry $g$ of a particular
string theory, it is always possible
to construct a background where one direction $x\uu 9$ is
compactified on a circle $S\uu 1$ of radius $R\ll 9$,
with a Wilson line around
that circle which implements the symmetry $g$ on string
states.
We will consider the case where $g$ is $(-1)\uu{F_{L_S}}$,
the operation which acts with a $-1$ on all spacetime fermions
coming from left-moving spin fields $\sst$
in the RNS formalism,
or left-moving worldsheet
fermions $\tht$ in the GS formalism.  $g$ acts with
a +1 on the corresponding right-moving fields.  Consistency
then demands that masless $p$-form potentials are also
odd under $g$, so they will be antiperiodic on the
circle $x\uu 9$.

If we had chosen $g$ instead to be $(-1)\uu{F_S}$, 
with $F\ll S$ the \it total \rm left- and right-moving
spacetime fermion number, we would have a conventional
Scherk-Schwarz compactification, such as
have been studied extensively in string
theory \cite{ssc}.  Our Wilson line can be thought of
as a kind of chiral Scherk-Schwarz theory.

\subsection{Closed strings in the Green-Schwarz formalism}

\heading{States}

In the GS formalism, we take
string states 
to have nonzero $p\ll -$, and choose
light-cone gauge.  
The gauge-fixed worldsheet theory has
eight transverse bosons $X\uu i$ with $i\in\{2,
\cdots ,9\}$ and eight fermions of each worldsheet chirality. 
Arbitrary 
string states with $p\ll - \neq 0$
correspond to level-matched states
of the gauge-fixed theory. 
Their chirality under the transverse $SO(8)$ is determined
by which type II theory we are dealing with.  We can always
take the left-movers $\tht\ll a$ to be in the ${\bf 8}\ll s$.
In type IIB the $\th\ll a$ are in the ${\bf 8}\ll s$ as well,
and in type IIA the $\th\ll \aad$ are in the ${\bf 8}\ll a$.
In addition to being
Weyl with some particular chirality, all our spinors are
also Majorana, so they comprise a total of eight real fermionic
fields on the left and on the right.

Unlike the worldsheet fermions of the RNS formalism,
the $\th,\tilde{\th}$ fermions are always periodic on the
cylinder.\footnote{Despite this, vertex operators for spacetime
fermions and $p$-forms do not give rise to cuts in $\th$.  This is
not a contradiction, due the lack of symmetry in the light-cone gauge
between states on the one hand, which 
carry $p\ll - \neq 0$, and operators on
the other hand, which carry
$p\ll - = 0$.  Vertex operators
for states with $p\ll - \neq 0$ have an involved expression
and we do not need to consider them.}
The mass-squared of the string state is just determined by
the energy of the state, in units of $\apr$:
\bbb
m\sqd = 2~E\ll{ws}~r\ll{ws} / \apr = {2\over{\apr}}
(L\ll 0 + \tilde{L}\ll 0 - 1).
\eee
Here $r\ll{ws}$ is the coordinate radius of the
worldsheet and $L\ll 0, \tilde{L}\ll 0$ are
Virasoro generators of the CFT of the gauge fixed
Green-Schwarz worldsheet.
With these rules, we calculate the lowest states
in the spectrum of the IIA
theory.  In the untwisted sector, the ground state energy is
zero.   The zero modes of the $\th$ fermions generate
an ${\bf 8}\ll s \oplus {\bf 8}\ll v$ and the zero modes
of the $\tilde{\th}$ fermions generate an
${\bf 8}\ll a \oplus {\bf 8}\ll v$.  The operator $g$ acts
as $g\ll{\tilde{\th}} \cdot g\ll {X\uu 9}$.  On states with
no oscillators excited, $g\ll{\tilde{\th}}$ acts with a
$+$ sign on the left-moving ${\bf 8}\ll v$ and a $-$ sign
on the left-moving ${\bf 8}\ll a$ states.  $g\ll {X\uu 9}$
acts only on the zero mode $x\ll 9$, as $x\ll 9 \to x\ll 9
+ 2\pi R\ll 9$, so that it acts as $\exp{2\pi i p\ll 9 R\ll 9}$,
which gives a + sign on states with $p\ll 9 R\ll 9
\in \IZ$ and a $-$ sign on states with $p\ll 9
R\ll 9\in \IZ + \hh$.  So the integral KK modes on the circle
live in the ${\bf 8}\ll v$ sector of the left-moving
Hilbert space, and the half-integral modes live in the
${\bf 8}\ll a$ of the left-moving Hilbert space.

Let us use the notation $n\ll 9$ for
$p\ll 9 R\ll 9$.
The untwisted spectrum with no oscillators excited contains
the following states:
\bi
\item{Bosonic states in the ${\bf 8}\ll v\up L \otimes {\bf 8}
\ll v\up R$ of SO(8) with $n\ll 9 \in \IZ$.  This sector
contains the graviton, dilaton, and NS-NS antisymmetric tensor.
These states are periodic around the $S\uu 1$.}
\item{Fermionic states in the ${\bf 8}\ll v\up L
 \otimes {\bf 8}
\ll s\up R$ of SO(8) with $n\ll 9 \in \IZ$.  This sector
contains the right-moving gravitini and dilatini, all periodic
around the $S\uu 1$.}
\item{Fermionic states in the ${\bf 8}\ll a \up L\otimes
{\bf 8}\ll v\up R $ of SO(8) with $n\ll 9 \in \IZ + \hh$.  This
sector contains the left-moving gravitini and dilatini.  These
states are antiperiodic around the $S\uu 1$.}
\item{Bosonic states in the ${\bf 8}\ll a\up L \otimes
{\bf 8}\ll s\up R$ of SO(8) with $n\ll 9 \in \IZ + \hh$.  This
sector contains $p$-form potentials of rank 1 and 3, and
they are antiperiodic around the $S\uu 1$.}
\ei
To each of these sectors we add states which are
identical except that they contain a winding
number $w\equiv \Delta x\uu 9 / 2\pi R\ll 9$ which is
even.  For $w\in 2\IZ$ the boundary conditions on
all worldsheet fields are exactly as in the sector of
$w = 0$.

Next we consider the twisted sector, which has winding number
$w\equiv \Delta x\ll 9 / 2\pi R\ll 9$ equal to $1$ mod 2.
In the sector with no KK momentum, the winding contributes
an energy of $\Delta E\ll {ws} ~r\ll{ws} = w\sqd
~R\ll 9\sqd / \apr$.  The $\tilde{\th}$ fermions are
antiperiodic in this sector, so they contribute
a ground state energy of $\Delta \tilde{L}\ll 0
= - {8\over{16}} = -\hh$.
There is no corresponding $L\ll 0$ contribution, so the
ground states are not level matched in this sector.
The level matched states are obtained in one
of two ways: by acting with
an odd number of $\tilde{\th}$ oscillators, since these
are half-integrally moded, or by adding a half-integral
value of $n\ll 9$, since fractional momentum contributes
to the level mismatch by $w~n\ll 9$.  We can get
a level-matched twisted sector by inferring that
$g$
has an anomalous phase of $-1$ in the sectors with
odd winding.\footnote{In the
worldsheet CFT of asymmetric Wilson lines, there is a
general prescription for assigning ground state phases
in winding sectors such that the worldsheet CFT is modular
invariant.  This prescription was alluded to in
\cite{Hellerman:2004zm}
and will be discussed in more detail in future work.}
So the true action of $g\ll{X\uu 9}$
should be $\exp{2\pi i n\ll 9 + \pi i w}$.

The oscillator ground states of
the twisted sectors contain the following:

\bi
\item{Bosonic states in the ${\bf 1}\up L
\otimes{\bf 8}\ll v\up R$
of SO(8) with $n\ll 9 w = -\hh$.}
\item{Fermionic states in the ${\bf 1}\up L
\otimes{\bf 8}\ll s\up R$
of SO(8) with $n\ll 9 w = - \hh$.}
\ei
We can also act with a $\tilde{\th}\ll{-\hh}$ oscillator,
to level match states with no KK momentum.  The $\tilde{\th}$
oscillators transform in the ${\bf 8}\ll s$ of SO(8).  Doing
this gives the following states:
\bi
\item{Bosonic states in the ${\bf 8}\ll s
\up L \otimes{\bf 8}\ll s\up R$
of SO(8) with $n\ll 9 = 0$.  This sector has eight-dimensional
masses $m\ll 8\sqd  = w\sqd ~R\ll 9\sqd / \apr\sqd $.}
\item{Fermionic states in the ${\bf 8}\ll s\up L
\otimes {\bf 8}\ll v\up R$
of SO(8) with $n\ll 9 = 0$.  These also have
eight-dimensional masses $m\ll 8\sqd  = w\sqd ~R\ll 9\sqd
  / \apr\sqd $.}
\ei
In the limit $R\ll 9 \to 0$ these last two sectors give us
light fermion and $p$-form degrees of freedom.  We shall
discuss a T-dual interpretation of the spectrum shortly.

\heading{Full spectrum}

Now we will list the full spectrum of states.  We will
also organize the ground states in each sector by
their SO(8) quantum numbers, although SO(8) is broken by
the compactification.  The We have:
\bi
\item{Sector {\bf A}: Bosonic states with $n\ll 9\in \IZ$
and $w\in 2\IZ$ and an even number
of $\tth$ operators $N\ll{\tilde{\th}}\in 2\IZ$ acting on
the ${\bf 8}\ll v\up L \otimes {\bf 8}\ll v\up R$ state of
the fermion zero modes.  The
$\tth$'s are integrally moded.}
\item{Sector {\bf B}: Fermionic states with $n\ll 9\in \IZ$
and $w\in 2\IZ$ and an even number
of
$\tth$ operators $N\ll{\tilde{\th}}\in 2\IZ$ acting on
the ${\bf 8}\ll v\up L \otimes {\bf 8}\ll s\up R$ state of
the fermion zero modes.  The
$\tth$'s are integrally moded.}
\item{Sector {\bf C}: Fermionic states with $n\ll 9\in \IZ + \hh$
and $w\in 2\IZ$ and an odd number
of
$\tth$ operators $N\ll{\tilde{\th}}\in 2\IZ + 1$ acting on
the ${\bf 8}\ll v\up L \otimes {\bf 8}\ll v\up R$ state of
the fermion zero modes.  The
$\tth$'s are integrally moded.}
\item{Sector {\bf D}: Bosonic states with $n\ll 9\in \IZ + \hh$
and $w\in 2\IZ$ and an odd number
of
$\tth$ operators $N\ll{\tilde{\th}}\in 2\IZ+1$ acting on
the ${\bf 8}\ll v\up L \otimes {\bf 8}\ll s\up R$ state of
the fermion zero modes.  The
$\tth$'s are integrally moded.}
\item{Sector {\bf E}:
  Fermionic states with $n\ll 9\in \IZ$
and $w\in 2\IZ+1$ and an odd number
of $\tth$ operators $N\ll{\tilde{\th}}\in 2\IZ+1$ acting on
the ${\bf 1}\up L \otimes {\bf 8}\ll v\up R$ state of
the fermion zero modes.  The
$\tth$'s are half-integrally moded.}
\item{Sector {\bf F}:
  Bosonic states with $n\ll 9\in \IZ$
and $w\in 2\IZ+1$ and an odd number
of $\tth$ operators $N\ll{\tilde{\th}}\in 2\IZ+1$ acting on
the ${\bf 1}\up L \otimes {\bf 8}\ll s\up R$ state of
the fermion zero modes.  The
$\tth$'s are half-integrally moded.}
\item{Sector {\bf G}:
  Bosonic states with $n\ll 9\in \IZ+\hh$
and $w\in 2\IZ+1$ and an even number
of $\tth$ operators $N\ll{\tilde{\th}}\in 2\IZ+1$ acting on
the ${\bf 1}\up L \otimes {\bf 8}\ll v\up R$ state of
the fermion zero modes.  The
$\tth$'s are half-integrally moded.}
\item{Sector {\bf H}:
  Fermionic states with $n\ll 9\in \IZ+\hh$
and $w\in 2\IZ+1$ and an even number
of $\tth$ operators $N\ll{\tilde{\th}}\in 2\IZ+1$ acting on
the ${\bf 1}\up L \otimes {\bf 8}\ll s\up R$ state of
the fermion zero modes.  The
$\tth$'s are half-integrally moded.}
\ei

In sectors ${\bf A} - {\bf F}$ the nine-dimensional
mass of the lowest state
is $m\sqd\ll 9 =  ({{|n\ll 9|}\over{R\ll 9}}
+ {{R\ll 9 |w|}\over{\apr}})\sqd  $.  Excited states
are obtained by adding $k$ units of
oscillator weight to both $L\ll 0$ and $\tilde{L}\ll 0$
with $k\in\{0,1,\cdots\}$.  In every
sector this gives a mass spectrum of $m\ll 9\sqd
= ({{|n\ll 9|}\over{R\ll 9}}
+ {{R\ll 9 |w|}\over{\apr}})\sqd + {{4k}\over{\apr}}$.

In sectors ${\bf G}$ and ${\bf H}$ the nine-dimensional
mass of the lowest state is $m\sqd\ll 9 =  ({{n\ll 9}\over{R\ll 9}}
+ {{R\ll 9 w}\over{\apr}})\sqd  $ if $n\ll 9 w \geq -\hh$,
and $m\sqd\ll 9 =  ({{n\ll 9}\over{R\ll 9}}
- {{R\ll 9 w}\over{\apr}})\sqd - {2\over{\apr}}  $
if $n\ll 9 w \leq - \hh$.  Note that these
two expressions agree when $n\ll 9 w = - \hh$.  So the
spectrum in sectors ${\bf G,H}$ is
$m\sqd\ll 9 =  ({{n\ll 9}\over{R\ll 9}}
+ {{R\ll 9 w}\over{\apr}})\sqd  + {{4k}\over{\apr}}$
if $n\ll 9 w \geq -\hh$ and $m\sqd\ll 9 =  ({{n\ll 9}\over{R\ll 9}}
- {{R\ll 9 w}\over{\apr}})\sqd + {{4k-2}\over{\apr}}  $
if $n\ll 9 w \leq - \hh$, for $k\in \{0,1,\cdots\}$.

In the limit $R\ll 9\to\infty$ the states in sectors
${\bf E}-{\bf H}$ have $w \neq 0$, so they
go to infinite mass.  The states with $w = 0$ and
fixed oscillator state form
continua labelled by $p\ll 9 = n\ll 9 / R\ll 9$.
These states live only in the sectors ${\bf A,B,C}$
and ${\bf D}$.  Sector ${\bf A}$ contains the fields
$G\ll{\m\n},B\ll{\m\n},$ and $\Phi$.  Sector ${\bf D}$
contains the fields $C\ll\m\up{{\rm RR}}$ and $C\ll{\m\n\s}\up
{{\rm RR}}$.  Sectors ${\bf B}$ and ${\bf C}$ respectively contain
gravitini and dilatini of the two chiralities, though only those
in sector ${\bf B}$ have massless modes in 9 dimensions for
finite $R\ll 9$, indicating that the Wilson line breaks
the spacetime SUSY down to 16 supercharges in 9 dimensions.

The action of SUSY is particularly transparent here.  Since
the spacetime supercharges are generated entirely from
right-moving currents, the spacetime supercharges
must exchange sectors ${\bf A}\leftrightarrow {\bf B}$,
sectors ${\bf C}\leftrightarrow {\bf D}$, sectors
${\bf E}\leftrightarrow{\bf F}$ and sectors
${\bf G}\leftrightarrow{\bf H}$.  As for the
multiplet structure for the massless states, we can infer
it by state counting alone.  There are 64 bosonic
and 64 fermionic massless degrees of freedom in sectors
${\bf A}$ and ${\bf B}$ respectively.  The bosons are
the $27$ states of the nine-dimensional graviton,
the $21$ states of the nine-dimensional NS-NS B-field,
a KK vector $G\ll{i 9}$ and a B-vector $B\ll{i 9}$,
with 7 polarizations each, and two scalars $\Phi$ and $G\ll{99}$.
The bosonic
content of the
gravitational multiplet of ${\cal N}=1$ SUSY in 8+1 dimensions
is the graviton, a rank-two antisymmetric tensor, a
vector and a single real scalar.  The remaining bosonic
states -- a vector and a real scalar -- are the bosonic content
of a massless vector multiplet of ${\cal N} = 1$ SUSY in 9D.

So the low-energy theory at a generic radius $R\ll 9$
contains only a gravitational multiplet and a massless
vector multiplet.  Both
multiplets arise from sectors ${\bf A}$ and ${\bf B}$.  At
a particular radius the massless spectrum will be enlarged
by the appearance of extra vector multiplets, which
make the continuous gauge symmetry nonabelian.

\heading{T-duality}

In the limit $R\ll 9\to 0$ the states in sectors
${\bf C,D,G,H}$ have $n\ll 9\neq 0$, so they become
infinitely heavy and leave the spectrum.  States with
$n\ll 9 = 0$ and fixed oscillator state
form continua labeled by $w R\ll 9 / \apr$.
Sector ${\bf A}$ again contains a massless graviton,
two scalars, and a rank-two tensor.  Sector ${\bf B}$
still contains a massless gravitino and dilatino, even for
finite $\tilde {R}\ll 9$.  Sectors ${\bf E}$ and ${\bf F}$
contain fermions and $p$-forms, respectively,
which are massive for nonzero $R\ll 9$
but form massless continua in the limit $R\ll 9\to 0$.

For small $R\ll 9$ it is easiest to understand the spectrum
in terms of a T-dual theory.  The spectrum of winding
and momentum states
is invariant if we define
\bbb
\tilde{n}\ll 9 \equiv - w / 2
\\\\
\tilde{w} \equiv - 2 n\ll 9
\\\\
\tilde{R}\ll 9 \equiv \apr / (2 R\ll 9)
\eee
which means
\bbb
p\ll L \to + p\ll L \llsk\llsk\llsk p\ll R \to - p\ll R
\eee
The sectors, then are exchanged as
\bbb
{\bf A}\leftrightarrow {\bf A}\llsk\llsk
{\bf B}\leftrightarrow {\bf B}
\\\\
{\bf C}\leftrightarrow{\bf E}\llsk\llsk
{\bf D}\leftrightarrow{\bf F}
\\\\
{\bf G} \leftrightarrow {\bf G}  \llsk\llsk
{\bf H} \leftrightarrow {\bf H}
\eee
It is convenient to switch to a $T$-dual set of local
operators,
using the change of variables $X\uu 9\ll L \to + X\uu 9\ll L$,
$X\uu 9\ll R \to - X\uu 9\ll R$.

Let us now get a better sense of how the restored $SO(8)$
invariance looks in the limit $\tilde{R}\ll 9 \to \infty$.
   To ensure that the
right-moving spacetime supercurrent has a form with
manifest $SO(8)$ invariance in the limit $\tilde{R}\ll 9
\equiv \apr / (2 R\ll 9) \to \infty$, we must also change
$\th$ to $\G\uu 9~\th$.  This changes the
$SO(8)$ representation of the $\th$'s from ${\bf 8}\ll a$
to ${\bf 8}\ll s$, which in turn changes the $SO(8)$
representation of their ground states from ${\bf 8}\ll v\up R
\oplus {\bf 8}\ll s\up R$ to $\tilde{{\bf 8}}\ll v\up R
 \oplus \tilde{{\bf 8}}\ll a\up R$.  (The tildes remind
us that the new representation is really under a different
$SO(8)$, which rotates the T-dual coordinate $\tilde{X}\uu 9$ with
the other $X$'s, rather than the original $X\uu 9$.)

So when $R\ll 9$ goes to $\infty$, the gravitino states lie in
an ${\bf 8}\ll v\up L \otimes {\bf 8} \ll s\up R$ in
sector ${\bf B}$ and an
${\bf 8}\ll a\up L \otimes {\bf 8} \ll v\up R$ in
sector ${\bf C}$.
When $R\ll 9$ is small, we make the
replacements ${\bf 8}\ll a\up R \to {\bf {\tilde{8}}}\ll s\up R$
and ${\bf 8}\ll a\up R \to {\bf {\tilde{8}}}\ll s\up R$.
The massless gravitini come from sectors ${\bf B}$ and
${\bf E}$.  The $SO(8)$ representation of the
massless states in sector ${\bf B}$ is
\bbb
{\bf 8}\ll v \up L \otimes {\bf 8}\ll s \up R
= {\bf 8}\ll v \up L \otimes {\bf \tilde{8}}\ll a \up R.
\eee
The massless states in sector ${\bf E}$
transform as
\bbb
{\bf 8}\ll s\up L \otimes
{\bf 8}\ll v\up R
= {\bf 8}\ll s\up L \otimes {\bf \tilde{8}}\ll v\up R .
\eee

So the $SO(8)$ representations in the T-dual theory
contain one gravitino of each chirality.  Therefore the
$R\ll 9\to 0$ limit of type IIA on an $S\uu 1$
with a Wilson line for
$(-1)\uu{F\ll{L_S}}$ is another ten-dimensional
theory of type IIA, not
a theory of type IIB.

An identical set of calculations and deductions shows that
the $R\ll 9\to 0$ limit of type IIB on an $S\uu 1$
with a Wilson line for
$(-1)\uu{F\ll{L_S}}$ is another ten-dimensional
theory of type IIB.

This is different from the situation
of compactification on an ordinary circle, where taking the
$R\ll 9\to 0$ limit exchanges type IIA with type IIB.
In contrast to type II on an untwisted circle,
type II on
a circle with a Wilson line for $(-1)\uu{F_{L_S}}$ should
have a special radius $R\ll 9$ where the physics is self-dual,
just as in the case of the bosonic string.

\heading{Enhanced gauge symmetry at the self-dual radius}

When $R\ll 9 = \sqrt{{\apr}\over 2}$ the theory is
self-dual.  In addition to the obvious discrete symmetry
under $w\to - 2 n\ll 9, n\ll 9 \to -\hh w$, there is an
enhanced continuous symmetry as well.  In GS formalism
the enhanced spin-1 currents are obscure and involve
twist fields for the $\th$ fermions, and we will not
write them.  Instead we will examine
the spectrum and note that there are extra massless vector
states which carry winding and momentum on the circle $S\uu 1$.

By virtue of the mass formulae it is clear that
sectors ${\bf C,D,E,F}$ can never contain states with
$m\ll 9\sqd = 0$.  For arbitrary values of $R\ll 9$
the sectors ${\bf A,B}$ always contain states with $m\ll 9 = 0$,
which are precisely those with $w = n\ll 9 = 0$.
The sectors ${\bf G,H}$ can contain massless states
if and only if $R\ll 9$ is at the self-dual radius $R\ll 9
= (\apr / 2)\uu \hh$, and then only for $ w = -2 n\ll 9 = \pm 1$.
For those values, the oscillator ground states
are level matched with $m\sqd = 0$.  They
transform in the ${\bf 8}\ll v$ of $SO(8)$, which
decomposes as a vector and a scalar under the
little group $SO(7)$ of a massless field in $8+1$ dimensions.
Acting with a $\th$ zero mode
$\th\ll{0}\uu a$
gives the eight massless
polarizations of a minimal spinor of $SO(7)$.  Together these
fill out the states of a massless vector multiplet of
the ${\cal N} = 1$ supersymmetric gauge theory in $D = 8+1$.

We get one such multiplet for each of the two possible
values $\pm 1$ for $w = -2 n\ll 9$,
which labels the charge $q\ll L\equiv
n\ll 9 - \hh w$ under $U(1)\ll L$.
The charge $q\ll R \equiv n\ll 9 +\hh w$ under $U(1)\ll R$
is zero for both states.
Consistency of the 9D effective field theory
demands that the gauge group determined by the
massless vectors be unitary.  We have seen that it has
dimension 3 and is nonabelian, so the gauge group coming from
the left-moving currents must be $SU(2)$.  We shall be able
to examine the
algebra of currents explicitly when we go to the RNS
formalism.

Again by checking the mass formulae we can see that these
are the only extra massless states we pick up for any
value of $R\ll 9$.  So the enhanced gauge symmetry at
the self-dual radius is $SU(2)\ll L \times U(1)\ll R$.
Since SUSY is unbroken and the vacuum energy is zero,
this tells us the supersymmetry cannot be gauged.  At any
rate, there is not enough SUSY to form a
non-singlet representation of $SU(2)$.
We can conclude from this that
the graviphoton must be the gauge vector of $U(1)\ll R$,
while the $SU(2)\ll L$ vectors must
lie in vector multiplets
under the ${\cal N} = 1$
spacetime SUSY.  Counting of states supports the
same conclusion.

\heading{Vertex operators}

The construction of vertex operators in the light-cone
GS description
is somewhat involved and in the most familiar treatment
\cite{Green:1987sp}
leans heavily on spacetime supersymmetry, which is
partially broken by our Wilson
line compactification.  This will not impede us, since we will not be
interested in the detailed form of
the vertex operators.  We will be interested only in
the boundary conditions imposed by the vertex operators
on the worldsheet fields $X\uu i, X\uu 9, \th, \tht$.
While we only
consider \it states \rm with nonvanishing $p\ll -$
in light-cone gauge, the simplest \it vertex
operators \rm to construct are exactly those with $p\ll - = 0$.
As in \cite{Green:1987sp}, those shall be the ones we consider.

We will also be aided by a simple principle for
understanding the meaning of the vertex operators of
type $V\ll {bos}\up L\otimes V\ll{bos}\up R$ -- just as
in tbe bosonic string, they can
be thought of as the operators defined
by worldsheet Lagrangian perturbations induced
by giving an expectation value to a certain string state.
This will help us find the vertex operators corresponding
to sectors ${\bf A}$ and ${\bf D}$.

\bi
\item{$V\ll{\bf A}$: Vertex operators for states of
type ${\bf A}$ (except with vanishing $p\ll -$)
are of the form $:\exp{i k\ll i X\uu i}:
E\ll{(n\ll 9, w)}~V\up L~V\up R$ with $n\ll 9 \in \IZ$ and
$w \in 2 \IZ$.
Here, $E\ll{(n\ll 9, w)}\equiv :\exp{i p\ll 9\up L X\ll L\uu 9
+ i p\ll 9\up R X\ll R\uu 9
}:$ with $p\ll 9\up L \equiv {{n\ll 9}\over{R\ll 9}}
- {{w R\ll 9}\over\apr}$ and
$p\ll 9\up R \equiv {{n\ll 9}\over{R\ll 9}}
+ {{w R\ll 9}\over\apr}$.  The internal pieces $V\up L$
and $V\up R$ can be taken to be ordinary bosonic
operators with $\Delta\up R - \Delta\up L = w n$.  The
simplest representative of this sector is
$w = n\ll 9 = 0$, $k\ll i = 0$,
$V\up R = \pp\ll + X\uu j$ and $V\up L = \pp\ll - X\uu i$.
In the vicinity of such an operator the bosonic fields are single
valued (they may have meromorphic singularities) and the fermionic
fields $\th,\tth$ are also single valued.  In fact, at the level
of boundary conditions for operators, $E\ll{(integer,even~integer)}$,
$:\exp{ik\ll i X\uu i}:$, $V\up L$ and $V\up R$ are all
irrelevant.  For purposes of boundary conditions on
local fields, we can just take the operator to be $1$.
We shall write $V\ll{\bf A}\in [1]$.
}
\item{$V\ll{\bf B}$:  One can obtain a state
in sector ${\bf B}$ by acting with a right-moving
fermion $\th$ on a state in sector
${\bf A}$.  So we have
${\bf B } \in [\th]$.}

\item{$V\ll{\bf C}$: One can obtain a state in sector ${\bf C}$
by acting on a state in sector ${\bf A}$ with a left-moving
fermion $\tht$ and an exponential $E\ll{(\hh,0)}\equiv
:\exp{\pi i X\uu 9 / R\ll 9}:$ with half a unit of KK momentum.
So $V\ll{\bf C}\in [E\ll{(1/2,0)}~\tht]$.}

\item{$V\ll{\bf D}$:  One can obtain a state
in sector ${\bf D}$ by acting with a right-moving
fermion $\th$ on a state in sector
${\bf C}$.  So we have
$V\ll{\bf D } \in  [E\ll{(1/2,0)}~\th~\tht]$.}

\item{$V\ll{\bf E}$: This sector gives rise to branch cuts
in left-moving fermions $\tht$, so it includes a
factor of a twist field for $T$.  More precisely, let $T$
be a ground state twist field, and $S$
be a first excited twist field, with $T~\tht \sim \zb\uu{+\hh}
S$.  Then $V\ll{\bf E} \in [E\ll{(0,1)}~S] =
[E\ll{(0,1)}~\tht~T]$.  The exponential $E\ll{(0,1)}$ is there
because this sector has integral KK momentun and odd winding
on the $X\uu 9$ circle.}

\item{$V\ll{\bf F}$: One can obtain a state in sector ${\bf F}$
by acting with a right-moving fermion on a state in sector ${\bf E}$.
So we have $V\ll{\bf F}\in [E\ll{(0,1)}~\th~S] =
[E\ll{(0,1)}~\th~\tht~T]$.  Remember that the $S$ and $T$
fields are only spin fields for the $\tht$ fermions and
not for the $\th$ fermions.}

\item{$V\ll{\bf G}$: One can obtain a state in
sector ${\bf G}$ by acting on a state in
sector ${\bf E}$ with a left-moving
fermion $\tht$ and an exponential $E\ll{(\hh,0)}$
with half a unit of KK momentum.
So $V\ll{\bf G}\in [E\ll{(\hh,1)}~T]$.}

\item{$V\ll{\bf H}$:  One can obtain a state
in sector ${\bf H}$ by acting with a right-moving
fermion $\th$ on a state in sector
${\bf G}$.  So we have
$V\ll{\bf H } \in  [E\ll{(\hh,1)}~\th~T]$.}
\ei

For our purposes, the only meaningful distinction between the
$T$ and $S$ fields is that the $T$'s have $(-1)\uu{F_{L_S}}
= +1$ and the $S$'s have $(-1)\uu{F_{L_S}}
= -1$.

We will only make use of these operators in order to
derive constraints on the set of allowed D-branes in the theory.

\subsection{D-branes and open strings in the Green-Schwarz formalism}

The GS formalism in light-cone gauge can in a simple way only handle
branes which are extended in one time and at least one spatial
direction.  So for now we will consider only Dp-branes,
with $p = 1,\cdots,9$.  We can also treat zerobranes and
D-instantons when we turn to the NSR description of our theories.
Since we can always T-dual along an even number of trivial
circles to return to our original theory, we need
only consider branes which are extended along 7 or 8
of the $X\uu i$ directions.  The odd-dimensional branes,
which are unstable at large $R\ll 9$,
do not have simple, manifestly $SO(8)$-covariant descriptions,
but they will appear as T-duals of even dimensional branes
in the limit $R\ll 9\to 0$.

\heading{D8-branes localized on $S\uu 1$}

The simplest case to consider will be that of the D8-brane.
Since we are in type IIA string theory we expect this
brane to be stable in the limit $R\ll 9\to \infty$.
The boundary conditions\footnote{For brevity
we use the notation '$\ttat\pp$' to mean 'at the
boundary'.}
for the bosons $X\uu i, X\uu 9$
are
\bbb
\pp\ll n X\uu i =  0,~i = 2,\cdots,8~\ttat\pp
\\\\
X\uu 9 = 0 ~{\rm mod}~2\pi R\ll 9~\ttat\pp
\eee
In the limit $R\ll 9\to\infty$ this is just the standard
D8-brane.  The boundary condition for the scalars and fermions is
\bbb
\pp\ll n X\uu i = 0 ~\ttat \pp
\llsk\llsk X\uu 9 = 0~\ttat\pp
\llsk\llsk\th\ll\a = \G\uu 9 \ll{\a\bed} \tht\ll\bed
~\ttat\pp
\eee
By acting on a timelike boundary with the vertex
operators $E\ll{(0,2)}$ in sector ${\bf A}$ we can see we also
need to include all boundary conditions
\bbb
\pp\ll n X\uu i = 0 ~\ttat \pp
\llsk\llsk X\uu 9 = 4\pi j R\ll 9~\ttat\pp
\llsk\llsk \th\ll\a = \G\uu 9 \ll{\a\bed} \tht\ll\bed
~\ttat\pp
\eee
for all $j\in \IZ$.
We can also act with the vertex operators $V\ll{\bf F}$,
which change the sign of $\tht$ and change the value of $X\uu 9$
by $2\pi(j+1) R\ll 9$.  This set of operators generates the
boundary conditions
\bbb
\pp\ll n X\uu i = 0 ~\ttat \pp
\llsk\llsk X\uu 9 = 2\pi(2j+1)  R\ll 9~\ttat\pp
\llsk\llsk \th\ll\a = -\G\uu 9 \ll{\a\bed} \tht\ll\bed
~\ttat\pp
\eee
These are all the classical boundary conditions for
dynamical fields forced by the action of the
eight closed string sectors.

So the total set of allowed boundary conditions is
\bbb
\pp\ll n X\uu i = 0 ~\ttat \pp
\llsk\llsk X\uu 9 = 2\pi j   R\ll 9~\ttat\pp
\llsk\llsk \th\ll\a =(-)\uu j\G\uu 9 \ll{\a\bed} \tht\ll\bed
~\ttat\pp
\eee
for any integer $j$.

The physically allowed open string states are then
of the form
\bbb
\kket w \equiv \sum\ll j  ~\kket{\bm
{ \ttat L ~\pp:~~~\th
= (-1)\uu{j }\G\uu 9 \tht ,~X\uu 9 = 2\pi j  R\ll 9 \cr
   \ttat R ~\pp:~~~
\th
= (-1)\uu{j+w}\G\uu 9 \tht ,~X\uu 9 = 2\pi (j+w) R\ll 9 } \em } ,
\eee
which have winding number $w$.
Now let us examine the spectrum in the possible
open string sectors.  The details of the spectrum depend on
the value of $w$ mod 2.

\heading{Open strings with even winding}

The sum over $j$ just implements the periodicity of $X\uu 9$,
meaning that the spectrum is the same for all $j$,
and we take $j = 0$.
When $w$ is even, the left boundary
$\s\uu 1 = 0$ and right boundary $\s\uu 1 = 2\pi r\ll{ws}$
have the
same boundary condition $\tht = \G\uu 9
\th$ for the $\th,\tht$ variables.
For even $w$ the mode
expansion in the open string sector
with winding $w$ is
\bbb
\th\ll\a =
  \sum\ll{K = -\infty}\uu\infty~b\ll{\a K}
\exp{ iK (\s\uu 1 + \s\uu 0) / (2 r\ll{ws})}
\\\\
\tht\ll\ald  =
\sum\ll{K = -\infty}\uu\infty ~\G\uu 9\ll{\ald\a}b\ll{\a K}
\exp{iK (-\s\uu 1 + \s\uu 0) / (2 r\ll{ws})}
\eee
The operators $b\ll {K\a} = C\ll{\a\b} b\dag\ll{-K \a}$
create excitations with energy ${K\over{2 r\ll{ws}}}$
for $K < 0$ and destroy them for $K > 0$.  The modes
$b\ll{ 0 \a}$ generate a Clifford algebra of which
the ground states form a representation.  Though $SO(8)$
is broken by the boundary conditions on the $\tht$ fermions,
it is convenient to organize the fermion ground states
as representations under the $SO(8)$ which acts on the
$\th$ fermions.  The $\th$ fermions transform in the ${\bf 8}\ll a$
of $SO(8)$, so the fermion ground states transform in
the representation ${\bf 8}\ll v \oplus {\bf 8}\ll s$.
Under the unbroken $SO(7)$ this decomposes as
${\bf 7}\ll v \oplus {\bf 1}\oplus {\bf 8}\ll{Maj.}$.

For $w=0$ the content of the low-energy spectrum is
a massless vector multiplet of ${\cal N} = 1$
gauge theory in 9 dimensions, just as on an ordinary
D8-brane in type IIA string theory.  In fact, it is
immediate that the spectrum all open string sectors
with even $w$ is identical to what it would be for
a D8-brane transverse a circle of radius $R\ll 9$
with no Wilson line.  This is because the only effect
of the Wilson line on the open string sectors is to change
the boundary conditions on the $\th,\tht$ in sectors with
odd winding.

\heading{Open strings with odd winding}

In sectors of odd winding, the boundary condition is
$\tht = \G\uu 9 \th$ on the left and $\tht = - \G\uu 9\th$
on the right.  So the mode expansion for the fermions is
\bbb
\th\ll\a =
  \sum\ll{K  \in \IZ + \hh }~b\ll{\a K}
\exp{ i K (\s\uu 1 + \s\uu 0) / (2 r\ll{ws})}
\\\\
\tht\ll\ald  =
\sum\ll{K \in \IZ + \hh }~\G\uu 9\ll{\ald\a}b\ll{\a K}
\exp{iK (-\s\uu 1 + \s\uu 0) / (2 r\ll{ws})}
\eee
Now the frequencies of the eight sets of fermionic oscillators are
${{2K+1}\over{4 r\ll{ws}}}$ with $K$ in $\IZ$.
The eight bosonic oscillators have frequencies ${K\over{2r\ll{ws}}}$
with $K\in \IZ$ as usual.  The total ground state energy contributed
by all the oscillators together is $- {1\over {4r\ll{ws}}}$.
So the mass spectrum for odd winding $w$ is
\bbb
m\sqd\ll 9 = {{w\sqd R\ll 9\sqd}\over{\apr\sqd}}
- {1\over{\apr}} + {K\uu{tot}\over{\apr}}
\eee
where $K\uu{tot}$ is the total (integer) number of units
of oscillator energy.  Moreover, $K\uu{tot}$ is
even for bosonic states and odd for fermionic states, so
we can see that the supersymmetric Bose-Fermi
degeneracy is broken in sectors of odd
winding.  In fact for $w$ odd and $ R\ll 9 < \sqrt{\apr}/ |w|$
the lowest state of the open string is a tachyon!

The first excited states
come from acting with $b\ll{\a -1}$ on the vacuum.
These states transform as an ${\bf 8}\ll a$ under the
SO(8) which rotates the $\th$'s,
and as an ${\bf 8}\ll{Maj}$ under the unbroken $SO(7)$.
The mass of these fermions is given by
$m\sqd = w\sqd R\ll 9\sqd / \apr\sqd$, so they become
light in the $R\ll 9\to 0$ limit.  We shall
now examine that limit.

\heading{T-dual of the D8-brane}

Starting
at large radius and lowering $R\ll 9$, the eightbrane
is stable until we reach $R\ll 9 = \sqrt{\apr}$, which
is $\sqrt{2}$ times the self-dual radius.
We can continue lowering $R\ll 9$
past the point of instability, even past
the self-dual radius.  Since our brane is localized
in the $X\uu 9$ direction, it should be extended in the
T-dual coordinate $\tilde{X}\uu 9$, so clearly
going beyond the T-dual radius gives us a different
brane, though the closed string background
is of the same type with which we started.

To get some idea of what the spectrum of the T-dual brane
might look like, take $R\ll 9$ to zero and
consider modes with fixed $\tilde{p}\ll 9
\equiv \tilde{n}\ll 9 / \tilde{R}\ll 9 = 2 \tilde{n}\ll 9 R\ll 9
/ \apr = w R\ll 9 / \apr$.  In this limit the winding modes
form a continuum, and the distinction between even and odd
winding disappears -- we have both sets of states at every
value of $\tilde{p}\ll 9\in \IR$.

From the  spectrum with $w\in 2\IZ$, we get a set of
massless bosons which transform
under the $SO(8)$ of the $\th$'s as ${\bf 8}\ll v$.
Therefore it must also transform as an ${\bf 8}\ll v$
under the new $SO(8)$ of the T-dual theory.
The massless fermions of the even winding sector
transform as an ${\bf 8}\ll s$ under $SO(8)\ll \th$,
so they transform as an $\tilde{{\bf 8}}\ll a$ under
the $SO(8)$ of the T-dual theory, by the same argument
we made in the section on T-duality of the closed string
theory.

The odd-winding sectors contribute a real tachyon with
mass $m\ll 9\sqd = p\ll 9\sqd - {1\over{\apr}}$, which
corresponds to a field with $m\ll{10}\sqd = - {1\over{\apr}}$,
which is a tachyon in the ten dimensional sense.
We saw that the
first excited states of the odd-winding sectors are
a set of fermions with $m\ll 9\sqd = p\ll 9\sqd$,
so in the ten-dimensional sense they are massless.  Under
$SO(8)\ll\th$ they transform in the ${\bf 8}\ll a$, so
under the SO(8) of the T-dual theory they
transform in the $\tilde{{\bf 8}}\ll s$.

We have found
that the spectrum which is massless in the ten-dimensional
sense on the T-dual side is given by a vector and a
set of fermions which transform under the little group
as $\tilde{\bf 8}\ll a \oplus \tilde{\bf 8}\ll s$, corresponding
to a non-chiral Majorana fermion of $SO(9,1)$.  There is
also a real open string tachyon of mass-squared $-{1\over{\apr}}$.
At the massless and tachyonic level, we have the same open
string spectrum as that of a non-BPS ninebrane in type IIA
in 10 dimensions.  So the natural conjecture is that the
$R\ll 9\to 0$ limit
of a charged D8-brane transverse to an $S\uu 1$ with a
Wilson line for $(-1)\uu{F_{L_S}}$ gives an uncharged,
non-BPS D9-brane in type IIA.

Some basic checks support this idea.  The simplest
is the agreement of the 8+1 dimensional energy densities of the
two kinds of effective eightbrane.  Under T-duality on a
circle,
the dilaton must always transform such that
\bbb
{{R\ll 9}\over{g\ll s\sqd}} = {{\tilde{R}\ll 9}\over{g\ll s\pr
\sqd}}.
\eee
Since our T-duality transformation is
\bbb
\tilde{R}\ll 9 = {{\apr}\over{2 R\ll 9}},
\eee
we have
\bbb
{{g\ll s\pr}\over{g\ll s}} = \sqrt{\apr\over 2} ~R\ll 9\uu{-1}
= \sqrt{2\over{\apr}} ~\tilde{R}\ll 9
\eee
The tension of a D8-brane at string coupling $g\ll s$
is given \cite{Polchinski:1998rr} by
\bbb
T\ll{\rm D8}\up{g\ll s} = {1\over{g\ll s~(2\pi)\uu 8~\apr\uu{9/2}}}
\eee
Using the relation we just derived, we can rewrite this
expression in terms of the T-dual coupling and radius:
\bbb
T\ll{\rm D8}\up{g\ll s} = \sqrt{2}~
{{\tilde{R}\ll 9}\over{g\ll s\pr~(2\pi)\uu 8~\apr\uu{5}}}
=
\sqrt{2}~
{{2\pi \tilde{R}\ll 9}\over{g\ll s\pr~(2\pi)\uu 9~\apr\uu{5}}}
\\\\
= \sqrt{2}~T\ll {\rm D9}\up{g\ll s \pr}~(2\pi \tilde{R}\ll 9)
\eee
We have defined the quantity $T\ll{\rm D9}\up{g\ll s \pr}$ to be
$1 / (g\ll s\pr~(2\pi)\uu 9~\apr\uu 5)$, which is the
tension of a standard BPS D9-brane in type IIB string theory,
at string coupling $g\ll s\pr$.  If this were
equal to the tension of a non-BPS D9-brane in type IIA
string theory our T-duality conjecture would be
falsified.  However it has been observed
\cite{Horava:1998jy} that the
tension $T\ll{\rm non-BPS~D9}$ of a non-BPS ninebrane at
a given string coupling is larger than that of a
BPS D9-brane by a factor of $\sqrt{2}$.  As a result we
find
\bbb
T\ll{\rm D8}\up {g\ll s} = \sqrt{2}~T\ll{\rm D9}\up{g\ll s \pr}
~(2\pi\tilde{R}\ll 9)
= T\ll{\rm non-BPS~D9}\up{g\ll s \pr}~(2\pi\tilde{R}\ll 9)
\\\\
  T\ll{\rm non-BPS~D9}\up{g\ll s \pr}~\tilde{V}\ll 9,
\eee
which agrees with our T-duality proposal for the D8-brane.
Note that this agreement depends on the
extra factor of $2$ modifying the usual T-duality relation
between $R\ll 9$ and $\tilde{R}\ll 9$.

In type IIA a non-BPS ninebrane preserves the
symmetry $(-1)\uu{F_{L_S}}$ and the real open string tachyon
on the ninebrane is odd under this symmetry.  In the
T-dual picture, then, when $\tilde{R}\ll 9$ is finite,
the open string tachyon should be antiperiodic
around the $\tilde{X}\uu 9$ direction because there is a
Wilson line for $(-1)\uu{F_{L_S}}$ in the dual picture
as well as the original picture.  So only tachyon modes
with half-integral momentum $\tilde{n}\ll 9\in \IZ + \hh$ should
exist.  Since $\tilde{n}\ll 9 = w / 2$,
this agrees perfectly with the fact that in the
original eightbrane picture, the tachyonic modes exist
only for odd winding $w \in 2 \IZ + 1$.

It would be good to check
the T-duality conjecture in more detail, in particular to check
directly that the full $SO(9,1)$ or even just $SO(8)$ is
restored in the T-dual theory in the limit $R\ll 9\to 0$.
We will soon study these open string theories in
the RNS framework, where
Lorentz invariance in all
directions is more manifest.

\heading{Endpoint of tachyon condensation}

Now we take a slight detour into a question to which
we can offer only a conjectural answer.  The question
is: what is the endpoint of open string tachyon condensation
for the D8-brane on a circle, when $R\ll 9 < \sqrt{\apr}$?
The
default hypothesis might be that tachyon condensation should
lead to the closed string vacuum, since there are no
absolutely stable, charged branes to which it is possible to decay.

We claim that the actual situation is more interesting,
with an endpoint for the instabilities which
depends on the value of $R\ll 9$:
\bi
\item{For $\sqrt{\apr} < R\ll 9$ the brane may be
absolutely stable.}
\item{For $\sqrt{\apr} / 2 < R\ll 9
<\sqrt{\apr}$ it is possible
that the brane may decay completely,
since we know of no stable branes of any kind
in this range.  In particular, when $R\ll 9 > \sqrt{\apr} /
\sqrt{2}$,
the D8-brane is the lightest of the
effective eightbrane in
9 dimensions which we 
study here.  Other than the closed string vacuum, there is
no obvious lighter state to which it can decay.  See the
comments below, however, for discussion of a subtlety
relating to discrete charge conservation.}

\item{When
$R\ll 9 < \sqrt{\apr} / \sqrt{2}$, the non-BPS D9-brane in
the original IIA theory, wrapped on the $S\uu 1$,
is lighter than the D8-brane.  It is possible that the
non-BPS ninebrane could be the endpoint of
the decay of the D8-brane, though since it has
its own instability this seems unlikely without fine-tuning
or a symmetry.}

\item{For $R\ll 9 < \sqrt{\apr}
/ 2$
the D9-brane wrapped on $S\uu 1$
in the original IIA theory
becomes perturbatively stable.  We would now like to argue
that for $R\ll 9 \sqrt{\apr} / 2$ the endpoint
of open string tachyon condensation on the D8-brane
is actually a non-BPS D9-brane wrapping the circle.}
\ei

To illustrate this last decay
scenario, go to the T-dual frame, where the D8-brane
becomes a $\widetilde{\rm D9}$-brane wrapping the
$\widetilde{\rm S}\uu 1$.  The lightest winding tachyon
in the original theory becomes the lightest tachyon KK mode
in the dual theory -- the tachyon mode with $\tilde{n}\ll 9
= \hh$.  The profile of this mode is $\sin(\tilde{x}\ll 9 / (2
\tilde{R}\ll 9))$.  Assuming $\tilde{R}\ll 9$ to be
macroscopically large, we can treat the $\widetilde{\rm D9}$-brane
as noncompact, and the profile for the tachyon as
linear, with a zero at the origin $\tilde{x}\uu 9 = 0$.
The endpoint of such inhomogenous tachyon condensation
is believed to be a charged D8-brane localized at the origin
$\tilde{x}\ll 9 = 0$ \cite{Horava:1998jy}.  Returning to the original
IIA theory, this becomes a non-BPS D9-brane wrapping the
circle.  Since the circle has very small radius $R\ll 9$, this
transition is allowed energetically.  And there are no
perturbative instabilities of the endpoint of
the condensation for $R\ll 9 < \hh \sqrt{\apr}$, so the
a small perturbation to
final stages of the condensation will not destroy the
ninebrane.

The range $ \sqrt{\apr} / 2 < R\ll 9
<\sqrt{\apr} $ may display a
more subtle behavior than
simple decay to the closed string vacuum, due to
a possible mechanism which would rely on
a $\IZ\ll 2$-valued brane charge.

To motivate the existence of such a charge,
consider the theory at large radius $R\ll 9$. 
There, the eightbrane has no perturbative
instability, but there are virtual processes
involving topologically nontrivial
Euclidean eightbranes which can change the D8
into an anti-D8.  For instance, the D8 can
move around the circle by an amount $2\pi R\ll 9$;
when it comes back to itself the sign of the RR
charge has been reversed, and brane charge has
been violated mod 2.  Alternately, a D8-$\overline{\rm D8}$
pair can nucleate from
the vacuum, with one travelling around the $x\uu 9$-circle
while the other stays put.  One ends up with
two D8's or two $\overline{\rm D8}$'s.  
For large $R\ll 9$, this type of process, with
an Euclidean eightbrane worldvolume winding the circle,
is the only type of process by which eightbrane charge
can change.

We can see, then, that the large-radius effective theory
has a $\IZ\ll 2$-valued conservation law for 
eightbrane charge.
If one goes a step further and assumes that
stringy effects preserve this discrete conservation law,
then there is a discrete charge which should be
conserved mod 2 in any quantum process.
The presence of winding tachyons is a stringy effect
not described in the effective theory of 
branes and gauge quanta at large radius; their presence
completely changes the mechanism which violates
$\IZ$-valued brane charge, relative
to the brane-loop mechanism of the 
large-radius effective field theory.
But if $\IZ\ll 2$-valued charge conservation
survives into the stringy r\'egime, then
the endpoint of D8 decay could not
be the closed string vacuum.  Since there are
no 'simple' branes with eight
extended dimensions which are lighter than the
original D8 in this range, the endpoint
would have to be a described
by an interacting boundary CFT, possibly corresponding to
an inhomogeneous phase of the unstable ninebrane.

\subsection{Closed strings in the Ramond-Neveu-Schwarz formalism}

Having described the spectrum in the GS
formalism, we will now give a
description of the same theory in the
RNS framework.  This will make Lorentz invariance
and T-duality more transparent.  It will also allow us
to discuss branes which are localized in all the
noncompact spatial directions.  Finally, the RNS formalism
facilitates the construction of vertex operators by
reducing it to the
state-operator correspondence in a two dimensional CFT.

\heading{Fast review of the RNS approach}

Our CFT is a free one, based on ten worldsheet fields
$X\uu\m$ and their left- and right-moving superpartners
$\pst\uu\m, \psi\uu\m$.  The theory has
a $(1,1)$ worldsheet supersymmetry which commutes with
the $SO(9,1)$ and exchanges the $X\uu\m$'s with their
superpartners in the usual way.  The $SO(9,1)$ is broken
only by the compactification of the $X\uu 9$ direction.

In addition there are left- and right-moving $\tilde{b},
{\tilde{c}}$ and $b,c$ ghosts and also superghosts
$\tilde{\b},\tilde{\g}$ and $\b,\g$.  On the
cylinder the only effect of the ghosts and
superghosts is to contribute $-\hh$ to the
ground state weight of NS sectors and $-{5\over 8}$
to the ground state weight of R sectors.  Physical
states are superconformal primaries of weight zero,
which for us will be equivalent to superconformal
matter primaries of weight $+\hh$ in NS sectors and
$+{5\over 8}$ in R sectors.

We will
always use the terms R and NS to label sectors strictly
according to the periodicity of
the supercurrent in a given sector, rather than to refer
to boundary conditions for some particular free fermion.
The
notion of 'natural periodicity' -- periodic for
bosons and antiperiodic for fermions --
simplifies the computation of ground state
weights in a given sector; real fields (bosonic or
fermionic) with their natural periodicity contribute
$0$ to the ground state weight, and real fields
with unnatural periodicity contribute $+{1\over{16}}$.

In RNS framework the consistency conditions of a string
theory are particularly transparent: one needs only
modular invariance, tadpole cancellation, and closure
and single-valuedness of the algebra of vertex operators.
Modular invariance will follow from level matching and
the inclusion of twisted sectors; tadpole cancellation is
a result of the ${\cal N} = 1$ spacetime supersymmetry.
We will see these two features of the theory in this
section, and we will defer the discussion of vertex operators
to the next section.  Taken together, these properties
will demonstrate
the consistency of our theory.

\heading{Closed string states}

Since we have labelled the sectors ${\bf A}$ through ${\bf H}$,
we will be able to describe the states sector by sector in
the RNS formalism.  At the level of states on the cylinder
this is mostly straightforward.  The only subtlety
is that the GSO projection differs from the usual one
in certain sectors carrying momentum and winding.
This is necessary to ensure level matching, and as we shall
see later it is also necessary in order
to make the algebra of vertex
operators close with single-valued coefficient functions in
the OPE.

The superghosts contribute a minus sign to the GSO
phases $(-1)\uu{F\ll{L_W}}$ and $(-1)\uu{F_{R_W}}$
in NS sectors.  (By $F_{{L,R}_W}$ we mean left- and right-moving
\it worldsheet \rm fermion number.)
So the GSO projection for the matter states in NS
sectors is always opposite that for the states in the
full theory with ghosts and superghosts.  In the
usual type IIA superstring the GSO projection in the
full theory is + in NS sectors, + for right-moving
R sectors and - for left-moving R sectors.  We shall
see that this is different in our theory -- the constraints
of level matching demand that modify
all left-moving
GSO projections by a sign of $(-1)\uu{ w}$.
For convenience we will sometimes ignore the ghosts and
refer directly to the GSO projection in the
matter sector, which for NS states is opposite that
in the full theory with ghosts.  We will try to
observe this difference by distinguishing between
the 'full GSO' and 'matter GSO', as well as between
'matter vertex operators' and full vertex operators 
with ghost and superghost dressing.

\bi
\item{Sector \bb A : These are the NS-NS states with integral
momentum and even winding: $n\ll 9 \in \IZ$ and $w \in 2 \IZ$.
The GSO projection is
the usual one; the full GSO is +/+.  Therefore the
matter GSO is --/--.  For
$n = w = 0$ the oscillator vacuum of the matter theory has
weight $(0,0)$, so the lowest states satisfying the GSO
projection are $\pst\uu\m \ll{-\hh} \psi\uu\n\ll{-\hh} \kket 0$.
These are weight $(+\hh, +\hh)$, so the momentum
dressing must have weight zero.  The polarizations
of the massless fields
are transverse and are defined up to a null-state redundancy,
so the representations under the $SO(7)$ little
group are $\Lambda\sqd {\bf 7} \oplus {\rm Sym}\sqd {\bf 7}
\oplus {\bf 7} \oplus {\bf 7} \oplus {\bf 1}$,
which can be viewed as the decomposition of the
SO(8) representation ${\bf 8}\ll v\otimes {\bf 8}\ll v$ after the
SO(8) is broken to SO(7) by the compactification.  Therefore
the states with $m\ll 9\sqd = 0$,
can be identified with the nine-dimensional
graviton, dilaton, B-field, KK and
B vectors, and the modulus $G\ll{99}$}
\item{Sector \bb B: These are the NS-R states with
$n\ll 9 \in \IZ$ and $w \in 2 \IZ$.
The matter GSO projection is --/+.  The states
in this sector are the superpartners of the states in
sector \bb A.  The massless states are the dimensional
reduction of one set of
the ten-dimensional dilatino $\psi\ll\ald \up R$
and gravitino $\Psi\uu\m\ll\a{}\up R$.
The little group representation of the
massless fermions is the decomposition of ${\bf 8}\ll v
\otimes \bb 8 \ll s$ under the umbroken SO(7).}
\item{Sector \bb C : These are the R-NS states
with $n\ll 9 \in \IZ + \hh, w \in 2\IZ$.
The full GSO is --/+
and the matter GSO is --/--.  This sector contains
other
set of ten-dimensional gravitini and
dilatini $\psi\ll\a\up L, \Psi\uu\m\ll\ald{}\up L$.  These
fields
do not give rise to nine-dimensional massless
fields because they are antiperiodic on the circle due to the Wilson
line.  Indeed, tsector \bb C has no states which have
$m\ll 9\sqd$.}
\item{Sector \bb D : These are the R-R states
with $n\ll 9 \in \IZ + \hh, w \in 2\IZ$.
The full GSO is
--/--.  These states are the
superpartners of states in sector \bb C.  The states with
$w = 0$ represent modes of the ten-dimensional
RR vector $C\ll\m$ and three-form $C\ll{\m\n\s}$
which are antiperiodic on the $S\uu 1$.  Note that
since the physical polarizations of the RR forms
are antiperiodic, it follows from SO(8,1) invariance
that the timelike components $C\ll 0$ and $C\ll{0ij}$ must
be as well.  This gives a simple way to understand how
branes such as the D8-brane can be unstable in this
compactification: the lagrange multiplier fields which
enforce charge conservation are zero modes of
the $p$-form gauge fields, and these are projected out
by the Wilson line.}
\item{Sector \bb E : These are the strings in the R/NS
sector
with integral momentum $n\ll 9\in \IZ$ and odd winding
$w\in 2 \IZ + 1$.  The full GSO is +/+ and the matter
GSO is +/--.  These fermionic states are massive for
any nonzero value of $R\ll 9 = \apr / 2 \tilde{R}\ll 9$,
but they contain a massless continuum in the limit $\tilde{R}\ll 9
\to 0$ which we interpret as a second set of
ten-dimensional gravitini and dilatini in the 10-dimensional
T-dual theory.  In the RNS approach the SO(9,1) in the
$\tilde{R}\ll 9\to \infty$ limit is transparent, because
T-duality is simply the change of variables $X\ll R\uu 9
\to -X\ll R\uu 9$ along with $\psi\uu 9\to - \psi\uu 9$.
Since the spinor representntations are generated by
the Clifford algebras of products of worldsheet fermion
zero modes, it follows that T-duality reverses the $SO(9,1)$
chirality of right-moving R sectors and preserves the
$SO(9,1)$ chirality of left-moving R sectors. }
\item{Sector \bb F : These are the strings in the R/R
sector
with integral momentum $n\ll 9\in \IZ$ and odd winding
$w\in 2 \IZ + 1$.  They are the superpartners of
states in sector \bb E.  The full GSO is +/+.
This set of states contains a tower which becomes a massless
continuum of RR fields in the limit $\tilde{R}\ll 9\to\infty$.
Since T-duality reverses the sign of the GSO projection in
right-moving R sectors, the GSO projection of this sector in
T-dual variables is +/--, so the RR forms of the dual IIA
theory are a 1-form and 3-form, just as in the
original IIA theory.}
\item{Sector \bb G : These are strings in the NS/NS sector with
half-integral momentum $n\ll 9 \in \IZ + \hh$ and
odd winding $w \in 2\IZ + 1$.  The level mismatch
due to the winding and momentum is $L\ll 0 - \tilde{L}\ll 0$
is $n\ll 9 w$ mod 1, and it equals $\hh$ in this sector.
It must be cancelled with oscillator energies, so the
GSO projection must be +/-- or --/+ in order to have
level-matched states.  Closure of the algebra of
vertex operators (discussed in a later section) shows that
the consistent choice is --/+ for the full
GSO, which gives +/-- for the matter GSO.  This choice
of GSO projection is a bit unusual, since it means that
there are physical NS/NS vertex operators containing odd numbers
of worldsheet fermions.  However there is no inconsistency,
since the half-integral spin of these
operators is offset by the contribution coming from the
exponential $:\exp{i k\ll L X\ll L + i k\ll R X\ll R}:$,
which also has half-integral spin.  This sector
contains only massive states at generic radii, but
at the self-dual radius $R\ll 9 = \sqrt{{\apr}\over 2}$
it contains two extra massless gauge fields and a
complex massless scalar.  The
gauge fields are the states $e\ll i \psi\uu i\ll{-\hh}
\kket{0;~n\ll 9 = \pm\hh, w = \mp 1}$ and the
massless scalars are the states $\psi\uu 9\ll{-\hh}
\kket{0;~n\ll 9 = \pm\hh, w = \mp 1}$.  These states are
the off-diagonal components of
the $SU(2)$ vector boson and Hermitean adjoint scalar
at the self-dual radius.  They are the bosonic fields
of the 8+1 dimensional ${\cal N} = 1$ vector multiplet.}
\item{Sector \bb H: These are the strings in the
NS/R sector with $n\ll 9 \in \IZ + \hh$ and
odd winding $w \in 2\IZ + 1$.  The full GSO projection
for these states is --/+ and the matter GSO is
+/+.  They are the superpartners
of the states in sector \bb G.  Generically massive,
at the self-dual radius they contain the off-diagonal
fermionic components of the $SU(2)$ vector multiplet
of 8+1 dimensional ${\cal N} = 1$ SUSY.}
\ei

\heading{Closed string vertex operators
for the $(-1)\uu{F_{L_S}}$ Wilson line in RNS formalism}

To find the
$-1$ or $-\hh$ picture vertex operator corresponding
to a given superconformal primary matter state,
simply apply the state-operator correspondence in the
matter theory.  Then to
find the $0$ picture matter operators for NS
sectors, act with the raising operator $G\ll{-\hh}$
(or $\tilde{G}\ll{-\hh}$ for left movers).  This
is equivalent to applying a picture-changing operator
as described in \cite{Friedan:1985ge}.
We will describe the properties of the matter pieces
of the vertex operators only; including the
ghost and superghost factors in various
pictures works as usual.  Letting the indices $M,N,\cdots$
run from $0$ to $8$ we list the sectors.

\bi
\item{$V\ll{\bb A}$: This sector has $(-1,-1)$ picture matter vertex
operators NS$\ll -$/NS$\ll -$ with $w,2n\ll 9 \in 2\IZ$.
For instance
$\pst\uu\m \psi\uu\n$ are the matter vertex
operators for the lowest massless states.  The
corresponding
$(0,0)$ picutre operators are NS$\ll +$/NS$\ll +$
states, for instance $\pp\ll - X\uu\m \pp\ll + X\uu\n$.
These correspond transparently to the 9-dimensional
metric $G\ll{M N}$, B-field $B\ll{MN}$, dilaton $\Phi$
KK vectors $G\ll{M 9}, B\ll{M 9}$ and modulus $G\ll{99}$.}
\item{$V\ll{\bb B}$: These $(-1,-\hh)$ picture matter operators
are in the $\nsm / \rpp$ sector with $w,2n\ll 9\in 2\IZ$.
The $(0,-\hh)$ picture matter vertex operators are in
the $\nsp / \rpp$ sector.  For instance,
the operators include the massless
9D gravitini/dilatini matter vertex operators, $\pst\uu\m \ss\ll\a$
in the $(-1,-\hh)$ picture, $\pp\ll - X\uu\m \ss\ll\a
+ o (k\ll\m)$ in
the $(0,-\hh)$ picture.}
\item{$V\ll{\bb C}$: $\rmm/\nsm$ or $\rmm/\nsp$ matter vertex
operators in the $(-\hh,-1)$ and $(-\hh,0)$ pictures,
respectively, with $n\ll 9 \in \IZ + \hh$ and $w\in 2\IZ$.
A typical matter
vertex operator is $:\exp{i k\ll M X\uu M}:
~E\ll{(\hh,0)}~\tilde{\ss}\ll{\dot{\a}}~(e\ll\m \psi\uu\m)$
in the $(-\hh,-1)$ picture.  This is a massive spin-3/2 field
in 9 dimensions.}
\item{$V\ll{\bb D}$: $\rmm/\rmm$ matter vertex
operators  with $n\ll 9 \in \IZ + \hh$ and $w\in 2\IZ$.
A typical matter vertex operator is $:\exp{i k\ll M X\uu M}:
~E\ll{(\hh,0)}~\tilde{\ss}\ll{\dot{\a}}~\ss\pr\ll \ald$.
This is a set of massive $p$-form fields
in 9 dimensions.  Taking the OPE with $\ss\ll\a$ implements
SUSY transformations between sectors $\bb C$ and $\bb D$.}
\item{$V\ll{\bb E}$: These fermionic strings
are R/NS sectors with integral momentum $n\ll 9$ and odd
winding $w$.  The $(-\hh,-1)$ picture matter vertex operators
have GSO projection +/- and the $(-\hh,0)$ picture
matter vertex operators have GSO projection +/+.  A typical
operator is $:\exp{i k\ll M X\uu M}:
~E\ll{(0,1)}~\tilde{\ss}\ll{a}\pr~(e\ll\m \psi\uu\m)$
in the $(-\hh,-1)$ picture.}
\item{$V\ll{\bb F}$: These bosonic string modes
are R/R sectors with integral momentum $n\ll 9$ and odd
winding $w$.  The matter vertex operators
have GSO projection +/+.  A typical
operator is $:\exp{i k\ll M X\uu M}:
~E\ll{(0,1)}~\tilde{\ss}\ll{a}\pr~\ss\ll\a$.}
\item{$V\ll{\bf G}$: The matter vertex
operators for these bosonic strings have
$n\ll 9 \in \IZ + \hh$ and $w\in \IZ$.  They are in
the $\nsp/\nsm$ sector in the $(-1,-1)$ picture, and
the  in the $\nsm/\nsp$ sector in the $(0,0)$
picture.  A typical $(-1,-1)$ picture
matter vertex operator is $:\exp{i k \ll M X\uu M} : E\ll{(\hh,-1)}
~\psi\uu\m$.  Note the peculiarity that this
NS/NS vertex operator has odd worldsheet fermion number.  Yet it
has zero spin on account of the spin contributed by
$E\ll{(\hh,-1)}$.}
\item{$V\ll{\bf H}$: The matter pieces of
the vertex
operators for these fermionic strings have
$n\ll 9 \in \IZ + \hh$ and $w\in \IZ$.  They are in
the $\nsp/\rpp$ sector in the $(-1,-\hh)$ picture, and
the  in the $\nsm/\rpp$ sector in the $(0,-\hh)$
picture.  A typical $(-1,-\hh)$
matter vertex operator is $:\exp{i k \ll M X\uu M} : E\ll{(\hh,-1)}
~\ss\ll\a$.}
\ei

We summarize our set of sectors in table (\ref{wilsecs}).
\btt
\begin{center}
\bta{|c|c|c|c|}
\hline
Sector & n mod 1 & w mod 2 & b.c. and full GSO \\
\hline\hline 
{\bf A} & 0 & 0 & $\nsp/\nsp$ \\
\hline
{\bf B} & 0 & 0 & $\nsp/\rpp$ \\
\hline
{\bf C} & $\hh$ & 0 & $\rmm/\nsp$ \\
\hline
{\bf D} & $\hh$ & 0 & $\rmm/\rpp$ \\
\hline
{\bf E} & $\hh$ & 1 & $\rpp/\nsp$ \\
\hline
{\bf F} & $\hh$ & 1 & $\rpp/\rpp$ \\
\hline
{\bf G} & 0 & 1 & $\nsm/\nsp$ \\
\hline
{\bf H} & 0 & 1 & $\nsm/\rpp$ \\
\hline
\eta
\end{center}
\caption{Closed string sectors in the background of a
Wilson line for $\mfls$.}
\label{wilsecs}
\ett
One can then check directly the closure
of the algebra of vertex operators, and
we display the multiplication table in table (\ref{wilsmult}).
\begin{table}
\begin{center}
\bta{|c||c|c|c|c|c|c|c|c|}
\hline
 & \bf A & \bf      B & \bf      C & \bf      D & \bf      E & \bf      F & \bf      G & \bf      H \\
\hline
\hline \bf  A & \bf A & \bf  B    & \bf  C    & \bf D     & \bf E     & \bf F     & \bf  G    & \bf H           \\
\hline \bf  B & \bf   B   & \bf A     & \bf  D    & \bf  C    & \bf F      & \bf  E    & \bf  H    & \bf  G      \\
\hline \bf  C & \bf   C   & \bf D     & \bf  A    & \bf  B    & \bf    G  & \bf   H   & \bf E     & \bf  F      \\
\hline \bf  D & \bf D     & \bf  C    & \bf   B   & \bf A     & \bf H     &  \bf G     & \bf F     & \bf  E      \\
\hline \bf  E & \bf  E    & \bf F     & \bf  G    & \bf  H    & \bf A     & \bf B     & \bf  C    & \bf   D     \\
\hline \bf  F & \bf  F    & \bf  E    & \bf   H   & \bf G     & \bf B     & \bf A     & \bf  D    & \bf   C     \\
\hline \bf  G & \bf  G    & \bf H     & \bf   E   & \bf  F    & \bf C     & \bf  D    & \bf A     & \bf  B      \\
\hline \bf  H & \bf   H   & \bf  G    & \bf  F    & \bf E     & \bf D     & \bf C     & \bf  B    & \bf   A     \\
\hline
\eta
\end{center}
\caption{Multiplication rules for closed string vertex operators
in the background of a Wilson line for $\mfls$.}
\label{wilsmult}
\end{table}
We display a separate
list, table (\ref{wilsspec}), summarizing the massless spectrum
for $R\ll 9\neq 0,\infty$.
\btt
\begin{center}
\bta{|c|c|c|c|}
\hline
Field & Vertex Op. & Sector & Massless in 9D? \\
\hline
$G\ll {\m\n}, \Phi$ &
$\tilde{c} c ~
\exp{-\pht-\phi + i k\ll P X\uu P} \pst\ll {(\m}\psi\ll {\n)}$ & A
& always
\\
$B\ll{\m\n}$ &
$\tilde{c} c ~
\exp{-\pht-\phi + i k\ll P X\uu P} \pst\ll {[\m}\psi\ll {\n]}$ & A
& always
\\
$\Psi\uu \m\ll\a$ &
$\tilde{c} c ~
\exp{-\pht-\phi/2 + i k\ll P X\uu P} \pst\uu \m \ss
\ll\a $ & B & always
\\
\hline
$\tilde{A}\uu{1}\ll M \pm i \tilde{A}\uu 2\ll M$ &
$\tilde{c} c ~
\exp{-\pht-\phi + i k\ll P X\uu P} E\ll{(\pm \hh,\mp 1)} \psi\uu M
 $ &  G & at $R\ll 9
= \sqrt{{\apr\over 2}}$ \\
$ \phi\uu 1
\pm i \phi\uu 2$ &
$\tilde{c} c ~
\exp{-\pht-\phi + i k\ll P X\uu P} E\ll{(\pm \hh,\mp 1)} \psi\uu 9
 $ &  G & at $R\ll 9
= \sqrt{{\apr\over 2}}$ \\
$\Upsilon\ll\a\uu 1  \pm i \Upsilon\ll\a\uu 2$ &
$\tilde{c} c ~
\exp{-\pht-\phi/2 + i k\ll P X\uu P}  
E\ll{(\pm \hh,\mp 1)} \ss\ll\a
$ &  H & at $R\ll 9
= \sqrt{{\apr\over 2}}$ \\
\hline
\hline
\end{tabular}
\end{center}
\caption{Massless sectors for finite $R\ll 9, \tilde{R}\ll 9$
with a Wilson line for $\mfls$.}
\label{wilsspec}
\ett

\heading{Open string vertex operators in the RNS framework}

The vertex operators for open strings
in the RNS framework
are similar to the ones for closed strings.
In introducing the vertex operators, it is useful to
define the boundary spin fields $\s\ll \a, \s\pr\ll\ald$.  
Ordinarily these are simply taken to be boundary
values of the bulk spin fields $\ss,\sst$.  Since
our branes do not preserve spacetime supersymmetry,
there can be no linear relation between the
operators $\ss$ and $\sst$ at the boundary which
holds as an operator
identity across all sectors.  Therefore it is worth
paying a bit more attention than usual to
the properties of the $\s\ll\a,\s\pr\ll\ald$.

We should define the $\s\ll\a,\s\pr\ll\ald$
to be the fields which change the boundary
conditions of worldsheet fermions from $\psi = \pm
\pst$ to $\psi = \mp \pst$.  To do this, the boundary
spin field must create an equal and opposite phase
in the fermions as they pass it on the right -- a phase
of $\exp{+\pi i / 2}$ for the $\psi$'s and $\exp{-\pi i
/ 2}$ for the $\pst$'s, as one moves from
$z \in - i \IR\uu +$ to $z\in + i\IR\uu +$, through
a contour with $\re~ z > 0$.  Therefore the boundary
spin fields should have OPE's such as:
\bbb
\s\ll\a (0) ~\psi\uu\m (z) \sim z\uu{-\hh} ~\G\uu\m\ll{\a\ald}~
\s\ll\ald\pr(0)
\\\\
\s\ll\a(0)~-c(\m)~\pst\uu\m(\zb)\sim \zb\uu{-\hh}~\G\uu\m\ll{\a\ald}~
\s\ll\ald\pr(0)
\eee
where $c(\m)$ is a constant depending on
the index $\m$.  For the eightbrane $c(\m) = +1$ 
for $m\neq 9$ and $-1$ for $\m = 9$ in NS sectors.
This choice enforces the boundary condition
$\psi\uu\m = c(\m) \pst\uu\m$ for $z$ on the
negative imaginary axis and $\psi\uu\m = -c(\m)\pst\uu\m$
for $z$ on the positive imaginary axis.
The boundary limit $\lim\ll{{\rm Re}~z \to 0}~\ss\ll\a$
of the bulk spin field has the correct
OPE with the $\psi$'s and $\pst$'s,
so we can take this as a definition:
\bbb
\s\ll\a\equiv \lim\ll{{\rm Re}~z \to 0}~\ss\ll\a
\eee
We define the $\s\pr\ll\ald$ similarly, in terms
of boundary limits of $\ss\pr\ll\ald$'s.
The operator $\lim\ll{{\rm Im} ~z\to 0} \G\uu 9\ll{\a\ald}
\sst\ll{\ald}$ has the same OPE's with the $\psi$'s
and $\pst$'s as does $\s\ll\a$.  This OPE is consistent with
the condition that $\ss\ll\a = \o~\G\uu 9\ll{\a\ald}
\sst\ll\ald $ for any value of $\o$.

It is not possible for $\o$ to be uniform over all sectors.
We know
this because there are vertex operators in the theory
with odd winding and a single $\pst$ fermion.
When such a vertex operator is transported past a boundary insertion
of a bulk spin field, there is a $-1$ in the presence of $\sst$ and
a $+1$ in the presence of a $\ss$.  At the same time, the
value of $X\uu 9 / 2\pi R\ll 9$ also changes by an odd number
at the point of insertion.  So the relative
sign between $\ss$ and $\sst$ at the boundary exactly
depends on the boundary value of $\exp{i X\ll i / 2 R\ll 9}$:
\bbb
\s\ll\a \equiv \lim\ll{{\rm Re}~z \to 0}~\ss\ll\a
= \exp{i X\ll i / 2 R\ll 9} ~\G\uu 9\ll{\a\ald} \sst\ll\ald
\eee
This relation implies that the spin fields 
$\ss,\sst$ have zero modes $\ss\ll 0 , \sst\ll 0$
on the
interval
exactly when the winding number, \it i.e. \rm $(X\ll 9 ~\ttat
~\rm L \it - X\ll 9 ~\ttat
~\rm R \it ) / {2\pi R\ll 9}$ is even.  This explains why the
spectrum looks supersymmetric in the even winding sectors but
nonsupersymmetric and potentially tachyonic in the
odd winding sectors.

Though the phase $\o$ varies from sector to sector, in
all sectors it is the case that the boundary
limits of the operator 
$\sst\ll\ald$ can be traded in for a boundary limit
of the $\G\uu 9\ll{\ald\a} \ss\ll\a$, with some coefficient.  
This establshes the consistency
of the multiplicative conservation,
between bulk and boundary, of the quantity 
$(-1)\uu{F_{L_W}}~\cdot ~(-1)\uu{F_{L_S}}$, the product
of left-moving worldsheet and spacetime fermion parities.  Though
$(-1)\uu{F_{L_W}}$ and
$(-1)\uu{F_{L_S}}$ 
cannot be conserved separately in the presence of a boundary,
the combination is preserved by our boundary conditions.  This
is precisely the same as in the case of ordinary $Dp$-branes
in type IIA with even $p$, but we emphasize it here
because the context is less familiar due to the more 
involved choice of sectors.  Below, we organize
products of bulk and boundary operators into a 
multiplication table, so that the conservation of
$(-1)\uu{F_{L_W}}~\cdot ~(-1)\uu{F_{L_S}}$ is easy
to check.  It is precisely this quantity which is
equal to $(-1)\uu w$ in all closed string sectors, and
it can therefore be transmitted to open strings via
the winding number.

Having defined the boundary spin fields $\s\ll\a$, we
can write the vertex operators for open strings in the
RNS formalism.
For the D8, we have the vertex operators are as follows
(ghost and $X\uu P$ dependence are omitted):
\bi
\item{Sector {\bf a}: $\nsp$ states with even winding $w\in 2\IZ$.
The matter GSO is $\nsm$.  This sector contains the gauge
field, whose vertex operator is $\lim \ll{{\rm Re} ~z\to 0}
\psi\uu M = \lim \ll{{\rm Re} ~z\to 0}\pst\uu M$.
The transverse scalar is also present, with vertex operator
$\lim \ll{{\rm Re} ~z\to 0}
\psi\uu 9 = - \lim \ll{{\rm Re} ~z\to 0} \pst\uu 9$.}
\item{Sector {\bf b}: $\rpp$ states with even winding $w\in 2\IZ$.
One such state is the massless fermion on the D8-brane,
whose vertex operator is $\s\ll\a$.}
\item{Sector {\bf c}: $\nsm$ states with odd winding $w\in 2\IZ
+ 1$.  The matter GSO is $\nsp$.  This sector contains the
tachyon, whose vertex operator is $\lim
\ll{{\rm Re} ~z\to 0} E\ll{(0,w)}~$, with
$w$ odd.}
\item{Sector {\bf d}: $\rmm$ states with odd winding $w\in 2\IZ
+ 1$.  This sector contains massive winding fermions
$\s\pr\ll\ald~\lim \ll{{\rm Re} ~z\to 0} E\ll{(0,w)}$
which become light only in the limit $R\ll 9\to\infty$.}
\ei

We summarize the open string sectors in table (\ref{sumopen}).
\btt
\begin{center}
\bta{|c|c|c|}
\hline
Sector & w mod 2 & b.c. and full GSO \\
\hline\hline 
{\bf a} & 0 & $\nsp$ \\
\hline
{\bf b} &  0 & $\rpp$ \\
\hline
{\bf c} & 1 & $\nsm$ \\
\hline
{\bf d}  & 1 & $\rmm$ \\
\hline
\eta
\caption{Open string sectors in the background of an eightbrane
in the Wilson line for $\mfls$.}
\label{sumopen}
\end{center}
\ett
The selection rules of the boundary OPE are summarized
by the multipication table (\ref{openmult}).
\btt
\begin{center}
\bta{|c||c|c|c|c|}
\hline
 & {\bf a} & {\bf b} & {\bf c} & {\bf d} \\
\hline\hline 
{\bf a} & \bf a & \bf b & \bf c & \bf d  \\
\hline
{\bf b} &  \bf b & \bf a & \bf d & \bf c \\
\hline
{\bf c} & \bf c & \bf d & \bf a  & \bf b \\
\hline
{\bf d}  & \bf d & \bf c & \bf b & \bf a \\
\hline
\eta
\caption{Multiplication rules for open string vertex operators
in the background of a D8-brane in the theory of
a Wilson line
for $\mfls$.}
\label{openmult}
\end{center}
\ett
There is also a consistent bulk-boundary OPE, as
demonstrated by the following multiplication 
rules for bulk operators on boundary operators
in table (\ref{openclosedmult}).

\btt
\begin{center}
\bta{|c||c|c|c|c|}
\hline
 & {\bf a} & {\bf b} & {\bf c} & {\bf d} \\
\hline\hline 
{\bf A, D} & \bf a & \bf b & \bf c & \bf d  \\
\hline
{\bf B, C} &  \bf b & \bf a & \bf d & \bf c \\
\hline
{\bf E, H}  & \bf d & \bf c & \bf b & \bf a \\
\hline
{\bf F, G} & \bf c & \bf d & \bf a & \bf b  \\
\hline
\eta
\end{center}
\caption{Multiplication rules for open string vertex operators
in the background of a D8-brane in the theory of
a Wilson line
for $\mfls$.}
\label{openclosedmult}
\ett

\subsection{Lift to M-theory}

It is in principle algorithmic to lift any Wilson
line background to M-theory.  A Wilson line
background is
simply string theory on a circle $x\uu 9$, a topologically
nontrivial space
patched together from locally trivial pieces, using
as a transition operation a discrete symmetry $g$
of type IIA string theory.  To find the corresponding
M-theory background, lift the discrete symmetry operation
to a symmetry $g\pr$ of 11 dimensional M-theory
on a circle $x\uu{11}$, and construct a circle
$x\uu 9$ patched together with the symmetry operation
$g\pr$.

In this case $g = \mfls$, which lifts to M-theory as
a reflection of $x\uu{11}$.  This means our Wilson
line background lifts to M-theory on a Klein bottle -- a
circle $x\uu{11}$ fibered over another circle $x\uu{9}$
which reverses orientation $x\uu{11}\to - x\uu{11}$ as 
it is transported around $x\uu 9$.

The lift illustrates previously unnoticed quantum properties
of M-theory.  In particular, the self-duality of
type IIA on the twisted circle lifts to a new duality
of M-theory which cannot be derived from any
previously known dualities.  In particular, though
M-theory is known to be self-dual on a $T\uu 3$, it is not
dual to itself when compactified on a circle, torus,
M\"obius strip or cylinder.

Let us now work out the duality in M-theory terms.
The parameters of the original M-theory are
\bbb
m\ll{pl} = g\ll s\uu{-1/3} \apr\uu{-1/2}
\\\\
R\ll{11} = g\ll s \apr\uu{+1/2}
\eee
Our duality changes the parameters by $R\ll 9 \to \apr / (2R\ll 9)$
and $g\ll s \to \sqrt{{\apr}\over 2} ~{{g\ll s}\over{R\ll 9}}$.
So the parameters of the dual M-theory compactification
are
\bbb
R\ll{11}\pr = \lrdd {{R\ll {11}}\over 2} \rrdd\uu{1/2}
m\ll p\uu{-3/2} R\ll 9\uu{-1}
\\\\
R\ll 9\pr = \hh (R\ll 9 R\ll{11} m\ll p\uu 3)\uu{-1}
\\\\
m\ll p\pr = (2 R\ll{11})\uu{1/6} R\ll 9\uu{1/3} m\ll p\uu{3/2}
\eee

As a check, note that the inverse
string tension $R\ll {11} m\ll p\uu 3$ and 
the nine-dimensional inverse Newton constant
$m\ll p\uu 9   R\ll 9 R\ll {11} $ are invariant
under this transformation.

The self-dual point is the point $R\ll 9 = \sqrt{\apr / 2}$. 
In eleven-dimensional terms this is when $R\ll {11}
 R\ll 9\sqd  = {1\over{2 ~ m\ll p\uu 3}}$.
In terms of massive charged states, this is the point
where a wrapped membrane has the same tension
as a mode with half a unit of Kaluza-Klein
momentum along the $R\ll 9$ direction.  At the self-dual radius
there is an enhanced gauge symmetry $SU(2)\times U(1)$,
as we saw in a previous section.  In the
eleven-dimensional picture, the W-bosons of the
enhanced gauge symmetry are membranes wrapping
the Klein bottle and carrying half a unit of $p\ll 9$.
The emergence of enhanced gauge symmetry from wrapped 
membranes in M-theory is not new altogether; 
a related effect plays an important role in type II-heterotic
duality, for instance \cite{Witten:1995ex}.
However this is a surprisingly
explicit example involving a completely flat background,
illustrating an ubiquitous theme of string
dynamics in a particularly
simple setting.

\section{An orbifold by reflection, with a $\mfls$ action}

The Wilson line for $(-1)\uu{F_{L_S}}$ is in some
sense a direct ingredient
in the nongeometric string
backgrounds of \cite{Hellerman:2002ax}.
In this section we shall examine
another type IIA
background which constitutes a second direct ingredient of
the theories described in \cite{Hellerman:2002ax}.

Our second ingredient is a $\IZ_2$ orbifold of four coordinates $x\uu{6-9}$, in which the orbifold action
has an extra phase of $(-1)^{F_{L_S}}$ relative to the usual orbifold action.
We shall see that this
extra phase changes drastically
the nature of the orbifold, altering the
massless content of the six-dimensional
theory on the singularity as well as the boundary
conditions for bulk fields at the origin.  The nature
of the supersymmetry of the 6D theory will be changed
relative to the usual orbifold.  While the theory
on an ordinary $\IR\uu 4 / \IZ\ll 2$ singularity has
$(1,1)$ supersymmetry in type IIA string theory, 
the theory on our modified singularity will have $(0,2)$
supersymmetry instead.  The properties of branes will also
be affected; the so-called 'regular' branes, which
are pointlike in the four transverse directions, have
instabilities sufficiently near the origin.  And the
'pinned' branes\footnote{In the standard orbifold
these would be called 'fractional'
branes.  However in our case we will see that they
do not carry any charge of
the bulk zerobrane, so the terminology 'fractional'
is inappropriate here.},
localized at the singularity, will have
an odd number, rather than even, of infinitely extended directions.

\subsection{Closed string states}

Now we will work out the closed string spectrum
of the $R\ll{6789}\cdot \mfls$ orbifold.
In the RNS formalism $(-1)^{F_{L_S}}$ acts only on spin fields for worldsheet fermions and not on ordinary worldsheet
fields themselves.  In the GS formalism, of course, the phase $(-1)^{F_{L_S}}$ acts on the
$\tht$ variables.  We could retrace the steps we took in the case of the Wilson line for
$(-1)^{F_{L_S}}$ and work out the spectrum in the GS formlaism
in light-cone gauge, later translating it to the
RNS formalism.  But having seen how this works in principle, we will not encumber the
reader with an elaborate working of the same steps in this case.  Instead we will proceed directly to
the RNS description.

\heading{Untwisted sector}

In the untwisted sector, the spectrum is simple to work out.  The worldsheet coordinates
$X\uu i$ with $i = 6,7,8,9$ have zero modes $x\uu i$ and the physical states on the
circle have a zero mode wavefunction
factor $f(x\uu i)$ which is an eigenfunction
under the Laplacian $-\sum\ll i {\pp\over{\pp x\uu {i2}}}$.  This Laplacian commutes with
the operator $R\ll{6789}$
which reflects $X\uu i$, so the physical states are common
eigenstates of
$R\ll{6789}$ and $k\ll i\uu 2$.
They are also separately eigenstates of $(-1)\uu{F_{L_S}}$, and the
orbifold projection in the untwisted sector simply demands that all states must have
the same eigenvalues under both operators.  Since one contribution to the eigenvalue of
$R_{6789}$ is determined by the change in sign of a wavefunciton upon reflection across the origin,  
the orbifold projection is
most simply expressed in terms of boundary conditions on bulk fields.

The effect of $R\ll{6789}
 \cdot (-1)\uu{F_{L_S}}$ is to give a phase $\pm$
with the following contributions:
\bi
\item{A $-$ sign for wavefunctions which are odd upon reflection and a $+$ sign for
wavefunctions which are even upon reflection.  This is equivalent to a $-$ sign for
modes which have 'D'-type boundary conditions at the origin, meaning that they vanish at $x\uu i = 0$,
and a $+$ sign for modes which have 'N'-type boundary conditions, meaning that their partial
derivatives $\pp\ll i f(x)$ vanish at $x\uu i  = 0$.}
\item{A $-$ sign for each Lorentz index oriented along the $x\uu i$ directions.}
\item{On spacetime fermions, a sign equal to the eigenvalue of $\G\uu{6789}$.  In type IIA
this is equal to the eigenvalue $\G\uu{012345}$ for right-moving fermions and $-\G\uu{012345}$
for left-handed fermions $\tilde{\Psi}$.}
\item{From the $(-1)\uu{F_{L_S}}$ piece, a sign of $(-1)$ for left-handed fermions and RR fields.}
\ei
Totether, the last two contributions combine to give a sign on all
fermions equal to the eigenvalue of $\G\uu{012345}$, in the case of type IIA.

The result is that the fields 
\bbb
G\ll{MN}, B\ll{MN}, G\ll{ij}, B\ll{ij}, \Phi, C\ll i, C\ll{iMN},
C\ll{ijk}, P\ll + \up 6\Psi\uu M, P\ll + \up 6\Pst\uu M, P\ll - \up 6 \Psi\uu i,
{~~~\rm and ~~~}
P\ll -\up 6\Pst\uu i
\eee
 have N-type boundary conditions, while the fields 
\bbb
G\ll{Mi}, B\ll{Mi}, C\ll M, C\ll{MNP},
C\ll{ijM}, P\ll - \up 6\Psi\uu M, P\ll - \up 6\Pst\uu M, P\ll + \up 6 \Psi\uu i,
 {\rm ~~~ and ~~~}
P\ll +\up 6\Pst\uu i
\eee
 have D-type boundary conditions.  (Here $M,N,\cdots$ run
from $0$ to $5$.)

Interpreted in terms of a choice of
untwisted sectors on the worldsheet, this set of boundary
conditions means that we have the usual orbifold
projection in the untwisted $\nsp/\nsp$ and $\nsp/\rpp$
sectors, and the \it opposite \rm of the usual orbifold
projection in the $\rmm/\nsp$ and $\rmm/\rpp$ sectors.
That is, in the $\rmm/\nsp$ and $\rmm/\rpp$ sectors
we keep states which are odd, rather than even, under
$R\ll{6789}$.

Bulk fields with N-type boundary conditions have zero modes
in six dimensions which are normalizable to the volume of
the transverse space.  Though these do not quite correspond
to fluctuating six-dimensional modes, this is an artifact
in some sense of the noncompactness of the local model.  The
zero modes of bulk fields with N-type boundary
conditions are \it almost \rm 6D fields,
since one would expect them to become dynamical in 6D once
the transverse space is compactified.  Fields with
D-type boundary conditions do not have zero modes
at all in six dimensions, even normalized to the volume
of the transverse space.

So in this sense, there are two six-dimensional
gravitini with the \it same \rm 6D chirality, corresponding
to (0,2) SUSY in the 6D effective theory.  This is
in contrast to the usual 6D effective theory in the
$\IR\uu 4 / \IZ\ll 2$ orbifold in type IIA, which has
(1,1) SUSY in six dimensions.  Given this, we would
expect that the normalizable modes in the 
6D theory would be tensor, rather than vector, multiplets. 
In the next section this expectation will be borne out.

\heading{Twisted sector} 

In the twisted sector the fields $X\uu i$ are antiperiodic on the string worldsheet, and
the zero modes $X\uu i\ll 0$ do not exist.  Their superpartners
$\pst\uu i, \psi\uu i$ have the periodicity opposite what
they would ordinarily have in the untwisted NS and R sectors.

Modular invariance under $\t\to - {1\over \t}$ demands that
we have exactly one twisted sector of
each type: NS/NS, NS/R, R/NS, R/R, with the orbifold
and chiral GSO projections determined by the requirement
of level matching in every sector.

First let us determine the correct orbifold projections.
In the twisted R/R sector, all worldsheet
fields $X\uu i ,\pst\uu i , \psi
\uu i$ which are odd under $R\ll{6789}$ are antiperiodic,
and $L\ll 0 - \tilde{L}\ll 0$ vanishes in the
oscillator ground state.  Therefore the orbifold
projection in the R/R sector must be $+$, for the
$-$ states all have $L\ll 0 - \tilde{L}\ll 0 \in \IZ + \hh$.

Taking the product of an even, twisted R/R state and
the untwisted states and imposing
closure of the algebra of vertex operators, we find that the 
twisted NS/NS and NS/R states must be -- under the orbifold 
projection; and twisted R/NS states must be +.

The possible GSO projections in the
twisted sector which are consistent with
closure of the vertex algebra can
be presented as NS$_\s$/NS$_{\s\pr}$, NS$_\s$/R$_{\s\pr}$,
R$_{-\s}$/NS$_{\s\pr}$, R$_{-\s\pr}$/R$_{\s\pr}$,
for some values of $\s\s\pr\in\{\pm 1\}$.
The consistent values of $\s,\s\pr$ can be 
derived from the requirements of level matching.
Since the NS/NS sector is odd under
$R\ll{6789}$, we must act on the vacuum
 with an odd number of the $\{\psi\uu i, \pst\uu i\}$.
This proves that $\s\s\pr = -1$, so the consistent
GSO must be of the form NS$_\s$/NS$_{-\s}$, NS$_\s$/R$_{-\s}$,
R$_{-\s}$/NS$_{-\s}$, R$_{-\s}$/R$_{-\s}$,
for some value of $\s\in\{\pm 1\}$.

So consider the sector R$_{-\s}$/NS$_{-\s}$, which
is R$_{-\s}$/NS$_{+\s}$ in the matter sector.
Including the superghost contribution, the ground
state level mismatch is zero mod 1.  The orbifold projection
is +, which means $\rnu\ll {X\uu i} + \rnu\ll{\psi\uu i}
+ \rnu\ll{\pst\uu i} \in 2\IZ$.   Level matching mod 1 means that
$\rnu\ll{\pst\uu i} + \rnu\ll{\psi\uu M} + \rnu\ll{X\uu i}
\in 2\IZ$.  Adding these two equations gives $\rnu\ll{\psi\uu i}
 + \rnu\ll{\psi\uu M} \in 2\IZ $ which means the 
matter GSO projection
on the right must be +, so $\s = +1$.

Now let us find the lowest-lying states
in each twisted sector.
\def\hh{1/2}
\bi
\item{Twisted $\nsp/\nsm$: The ground state weight
in the matter sector is $(1/4,1/4)$ from the $X\uu i$
and $(1/4,1/4)$ from the $\psi\uu i, \pst\uu i$.  So the
total weight is $(\hh,\hh)$, the correct weight
for a massless physical state in the NS/NS sector.  
All the $\pst\uu M, \psi\uu M$ oscillators have
nonzero energy, so the massless states must be
scalars in 6 dimensions.  The fermion zero modes $\pst\uu i\ll 0,
\psi\uu i\ll 0$ generate a Clifford algebra represented by
the ground states.  Under the $SO(4) \simeq SU(2) \times 
SU(2)$ group of rotations of the $X\uu i$, the left-moving
fermion ground states transform as $(2,1)$ and the
right-moving fermion ground states transform as $(1,2)$.  So
the total set of ground states transforms as $(2,2)$ 
under $SU(2) \times SU(2)$, which is the ${\bf 4}$ of SO(4).
We denote the massless twisted scalars by ${\cal M}\uu i$.
Immediately it is clear that the ${\cal M}\uu i$
cannot
be interpreted as geometric resolution moduli 
and B-field moduli through a collapsed two-cycle; geometric
resolution and B-field moduli
would transform in the $(1,3)$ and $(1,1)$ respectively
under the $SU(2)\times SU(2)$ transverse rotations.
The physical interpretation is interesting and puzzling.
Does conformal perturbation theory permit the ${\cal M}\uu i$
to be given a finite expectation value?  Given that they
are massless and that the spacetime theory has sixteen
supercharges, we expect that the answer must be affirmative.
What is the physical interpretation of the
theory with a finite value for ${\cal M}$?
We cannot answer that directly at this point.  Later
we shall consider type IIB on this background, S-duality
will give us some insight into the nature
of the resolved space.  We will see that
there is nonzero H-flux turned on when the
resolution parameter is nonzero.
As noted
earlier, the orbifold projection is -, but this imposes
no additional constraints beyond level matching.}
\item{Twisted $\nsp/\rmm$: the weight of the oscillator
ground state is $(\hh,5/8)$ in the matter sector, so the
lowest states are massless.  The $\pst\uu i$ are
periodic with GSO projection +, and the $\psi\uu i$
are antiperiodic, so the SO(4) representation
of the spacetime fermions is (2,1).  The $\psi\uu M$ are
periodic with GSO projection $-$, so the spacetime
fermions transform as
a SO(5,1) Weyl spinor with chirality --. 
Then level matched
states satisfying the GSO projection all automatically
survive the orbifold projection, which is opposite the
usual one: they
live in the subspace which has
eigenvalue $-$ under $R\ll{6789}$. 
We denote the massless states by $\l\ll\pd\uu\aald$.  Since the
fermion zero modes are real, a reality condition can
be imposed on this sector; the only such covariant condition
is that the spinor be pseudo-Majorana $\bar{\l}\uu\aald\ll p
= C\st\ll{p\pd}\e\uu{\aald \dot{B}} \l\uu{\dot{B}}\ll\pd$. }
\item{Twisted $\rmm/\nsm$: the weight of the oscillator
ground state is $(5/8,\hh)$ in the matter sector, so the
lowest states are massless.  The $\psi\uu i$ are
periodic with GSO projection --, and the $\pst\uu i$
are antiperiodic, so the SO(4) representation
of the spacetime fermions is (1,2).  The $\pst\uu M$ are
periodic with GSO projection $-$, so the spacetime
fermions transform as
a SO(5,1) Weyl spinor with chirality $-$.  The orbifold
projection is onto even states, and it is satisfied automatically
for level matched states which satisfy the GSO projection.
The spinor $\l\uu A\ll\pd$ is pseudo-Majorana $\bar{\l}\uu A
\ll p = C\st\ll{p\pd}\e\uu{AB}\l\uu B\ll\pd$.}
\item{Twisted $\rmm/\rmm$: The ground state
weight of the matter theory is $(5/8,5/8)$, so the lowest 
states are massless.  The $\pst\uu i, \psi\uu i$ are
antiperiodic, which means the $SO(4)$ representation
of these states is trivial.  The $\pst\uu M$ and $\psi\uu M$
are periodic, so the RR fields transform in the
chiral bispinor of SO(5,1) with both spinors
having the same chirality.  There are then four physical
states at each momentum,
corresponding to a RR scalar $A$ and a two-form
$A\ll{MN}$ 
with self-dual field strength in six dimensions
$dA\ll{MNP} = {1\over 6}
\e\ll{MNP}{}\uu{QRS}dA\ll{QRS}$.  The orbifold
projection onto even states is automatic in this
sector, given level matching and the GSO projection.}
\ei
Altogether the massless bosonic field content is a two-form with
self-dual field strength, and five real scalars -- exactly
the bosonic content of a $(0,2)$ tensor multiplet in
six dimensions. 
Note that the $SO(4)\simeq SU(2)\times SU(2)$
is an R-symmetry of the (0,2) SUSY
in 6D.  One SU(2) factor is the obvious SU(2) rotating
the two Weyl spinors, and the second SU(2) is the
quaternionic SU(2) performing phase
rotations on a Weyl spinor in 6D, and rotating it
into its conjugate.
\def\hh{{1\over 2}}
Once again we summarize the sectors in a
table:

\begin{center}
\bta{|c|c|c|c|c|}
\hline
Sector & $\rm{\bm{\rm eigenvalue
 \cr
\rm ~of ~\it R_{6789}}\em}$ & b.c.~ for~
 $X\uu i$ &
${\bm{\rm b.c. ~for \cr \tilde{G}/G  ~\rm and
\cr \rm full~ GSO}\em}$ & ${\bm{\rm massless \cr \rm content}\em}$
  \\
\hline\hline 
{\bf AA} & $+$ & untwisted & $\nsp/\nsp$ & $G\ll{\m\n},B\ll{\m\n}
,\Phi$ \\
\hline
{\bf BB} & $+$ & untwisted & $\nsp/\rpp$ & $\Psi\uu \m\ll\a$ \\
\hline
{\bf CC} & $-$ & untwisted & $\rmm/\nsp$ & $\Pst\uu \m\ll\aald$ \\
\hline
{\bf DD} & $-$ & untwisted & $\rmm/\rpp$ & $C\ll\m, C\ll{\m\n\s}$ \\
\hline
{\bf EE} & $-$ & twisted & $\nsp/\nsm$ &  ${\cal M}\uu i$\\
\hline
{\bf FF} & $-$ & twisted & $\nsp/\rmm$ & $\l\uu\aald\ll\pd $\\
\hline
{\bf GG} & $+$ & twisted & $\rmm/\nsm$ & $\l\uu A\ll\pd$ \\
\hline
{\bf HH} & $+$ & twisted & $\rmm/\rmm$ & $A\ll{MN}$ \\
\hline
\eta
\end{center}

\subsection{Vertex operators}

The construction of the vertex
operators is straightforward.  We present
the forms of the massless vertex operators
in the table below, using the following
notation:
\bi
\item{$\t$ is a ground-state twist operator for $X\uu i$}
\item{$\ss\ll A, \ss\pr\ll\aald$ are spin fields of
each $SO(4)$ chirality
for the fermions $\psi\uu i$.  $\sst\pr\ll A$ and $\sst\ll\aald$
are spin fields for the fermions $\pst\uu i$.}
\item{$\ss\ll p, \ss\pr\ll\pd$ are spin fields of
each $SO(5,1)$ chirality for the fermions $\psi\uu M$.
$\sst\ll p, \sst\pr\ll\pd$ are spin fields of
each $SO(5,1)$ chirality for the fermions $\pst\uu M$.}
\ei

\btt
\begin{center}
\bta{|c|c|c|c|}
\hline
field & matter vertex op. & sector  & at $X\uu i = 0$\\
\hline
$G\ll {MN}, \Phi, B\ll{MN} $ &
$ f\ll +(X\uu i)~
 \pst\ll M \psi\ll N$ & $\nsp/\nsp~\up +$ 
& N \\
$G\ll {ij}, B\ll{ij} $ &
$f\ll +(X\uu i)~
\pst\ll i \psi\ll j$ &$\nsp/\nsp~\up +$  
& N
\\
$G\ll {iI}+ B\ll{iI} $ &
$ f\ll -(X\uu i)~
 \pst\ll i \psi\ll I$ & $\nsp/\nsp~\up +$ 
& D
\\
$G\ll {iI}-  B\ll{iI} $ &
$ f\ll -(X\uu i)~
 \pst\ll I \psi\ll i$ & $\nsp/\nsp~\up +$ 
& D
\\
\hline
$P\ll 6 \up + \Psi\uu M$ &
$ f\ll +(X\uu i)~
\pst\uu M \ss\ll p \ss\ll A $ &  
$\nsp/\rpp~\up +$
  & N
\\
$P\ll 6 \up - \Psi\uu M$ &
$ f\ll -(X\uu i)~
\pst\uu M \ss\pr\ll \pd \ss\pr\ll\aald $ &  
$\nsp/\rpp~\up +$
  & D
\\
$P\ll 6 \up - \Psi\uu i$ &
$
 f\ll +(X\uu i)~
\pst\uu i \ss\pr\ll \pd \ss\pr\ll\aald $ &  
$\nsp/\rpp~\up +$
  & N
\\
$P\ll 6 \up + \Psi\uu i$ &
$
 f\ll -(X\uu i)~
\pst\uu i \ss\ll p \ss\ll A  $ &  
$\nsp/\rpp~\up +$
  & D
\\
\hline
$P\ll 6 \up + \Pst\uu M $ &
$
 f\ll +(X\uu i)~
\psi\uu M  \sst\pr\ll A \sst\pr\ll p $ &  
$\rmm/\nsp~\up -$
  & N
\\
$P\ll 6 \up - \Pst\uu M$ &
$
 f\ll -(X\uu i)~
\psi\uu M  \sst\ll\aald \sst\ll\pd $ &  
$\rmm/\nsp~\up -$
  & D
\\
$P\ll 6 \up - \Pst\uu i$ &
$
 f\ll +(X\uu i)~
\psi\uu i \sst\ll\aald \sst\ll\pd $ &  
$\rmm/\nsp~\up -$
  & N
\\
$P\ll 6 \up + \Pst\uu i$ &
$
 f\ll -(X\uu i)~
\psi\uu i  \sst\pr\ll A \sst\pr\ll p $ &  
$\rmm/\nsp~\up -$
  & D \\
\hline 
${}^{{}^{{}^{{}^{{}^{}}}}} dC\ll {MN}$ & $ 
 f\ll - (X\uu i) ~\sstb   \G\ll {MN} \ss$
 & $\rmm/\rpp ~\up - $  & D \\
$dC\ll {ij}$ & $ 
 f\ll - (X\uu i) ~\sstb  \G\ll {ij} \ss$
 & $\rmm/\rpp ~\up - $  & D \\
$dC\ll {iM}$ & $ 
 f\ll + (X\uu i) ~\sstb  \G\ll {iM} \ss$
 & $\rmm/\rpp ~\up - $  & N \\
$dC\ll{MNPQ}$ &
$f\ll - (X\uu i) ~\sstb  \G\ll {MNPQ} \ss$
 & $\rmm/\rpp ~\up - $  & D \\
$dC\ll{ij MN}$ &
$f\ll - (X\uu i) ~\sstb  \G\ll {ij MN} \ss$
 & $\rmm/\rpp ~\up - $  & D \\
$dC\ll{ij kl}$ &
$f\ll - (X\uu i) ~\sstb  \G\ll {ij kl} \ss$
 & $\rmm/\rpp ~\up - $  & D \\
$dC\ll{iMNP}$ &
$f\ll + (X\uu i) ~\sstb  \G\ll {iMNP} \ss$
 & $\rmm/\rpp ~\up - $  & N \\
$dC\ll{ij kM}$ &
$f\ll + (X\uu i) ~\sstb  \G\ll {ij kM} \ss$
 & $\rmm/\rpp ~\up - $  & N \\
\hline
${\cal M}\uu i$ & $\t~\sst \uu{\prime \dagger}
 \ll A  \G\uu i \ll{A\aald} 
\ss\pr \ll \aald$ & $\nsp/\nsm~\up -$ & localized \\
\hline
$\l\uu \aald \ll\pd$ & $\t~\sst\pr\ll A \ss\pr\ll\pd$ & 
$\nsp/\rmm ~\up - $ & localized \\ 
\hline
$\l\uu  A \ll\pd$ & $\t~
~\sst\pr\ll \pd
~\ss\ll A $ & 
$\rmm/\nsm ~\up + $ & localized \\ 
\hline
$d A
\ll {MNP} $ & $\t~{\sst\dag\ll\pd}  (\G\ll{MNP})\ll{\pd\dot{q}}
\ss\pr\ll{\dot{q}}$ & $\rmm/\rmm~ \up + $ & localized \\
\hline
\hline
\eta
\end{center}
\caption{Vertex operators for closed strings
in the orbifold by $R\ll{6789}$ with an
action of $\mfls$.}
\ett

Every vertex operator has an additional
factor of $\tilde{c} ~c~\exp{- \tilde{a}\tilde{\phi} - a\phi}
~ \exp{i k\ll P X\uu P}$, where $a,\tilde{a}$ are
equal to $1$ for NS sectors and $\hh$ for R sectors.

Checking the closure of the algebra of vertex operators
is straightforward, keeping in mind:
\bi
\item{Two twist operators $\t$ close on untwisted operators
involving the $X\uu i$.}
\item{Two $\ss\ll A$ or two $\ss\pr\ll\aald$ close on
untwisted operators involving the $\psi\uu i$, of even fermion
number, since they are spin fields for an even number
of complex fermions.  Likewise, two
$\sst\pr\ll A$ or two $\sst\ll\aald$ close on
operators involving the $\pst\uu i$, of even fermion
number.}
\item{A $\ss\ll A$ and a $\ss\pr\ll\aald$ close on
an untwisted operator involving the $\psi$, with
odd fermion number.  Likewise 
a $\sst\pr\ll A$ and a $\sst\ll\aald$ close on
an untwisted operator involving the $\pst$, with
odd fermion number.}
\item{Two $\ss\ll p$ or two $\ss\pr\ll\pd$ close on
untwisted operators involving the $\psi\uu i$, of odd fermion
number, since they are spin fields for an odd number
of complex fermions.  Likewise, two
$\sst\ll p$ or two $\sst\pr\ll\pd$ close on
operators involving the $\pst\uu i$, of odd fermion
number.}
\item{A $\ss\ll p$ and a $\ss\pr\ll\pd$ close on
an untwisted operator involving the $\psi$, with
even fermion number.  Likewise 
a $\sst\ll p$ and a $\sst\pr\ll\pd$ close on
an untwisted operator involving the $\pst$, with
even fermion number.}
\ei
One can check these multiplication rules, if necessary,
by grouping the real fermions arbitrarily
in pairs into complex fermions, and performing a
bosonization. 

\btt
\begin{center}
\bta{|c||c|c|c|c|c|c|c|c|}
\hline
\rown{}{A}{B}{C}{D}{E}{F}{G}{H} \\
\hline\hline
\rown A A B C D E F G H \\
\hline
\rown B B A D C F E H G \\
\hline
\rown C C D A B G H E F \\
\hline
\rown D D C B A H G F E \\
\hline
\rown E E F G H A B C D \\
\hline
\rown F F E H G B A D C \\
\hline
\rown G G H E F C D A B \\
\hline
\rown H H G F E D C B A \\
\hline
\eta
\caption{
Closure of the OPE in type IIA on 
$\IR\uu{5,1} \times \IR\uu 4 / \IZ\ll 2$ with an
action of $(-1)\uu{F_{L_S}}$.}
\end{center}
\ett

\subsection{Type IIB version and S-duality}

The type IIB version of this background is quite similar
at weak coupling to the type IIA version.  We shall describe it
briefly, with some comments at the end about
its S-dual version\footnote{See
also \cite{sens}, \cite{oren} for earlier work on the
type IIB theory on the $\mfls \cdot R\ll{6789}$ orbifold,
its S-dual, and its stable, non-BPS branes.}.

\heading{Bulk and localized spectrum}

Type IIB string theory on the $R\ll{6789}~\mfls$ 
orbifold can be obtained by 
starting with the type IIA theory on the same
orbifold background, and
performing a T-duality along one of the spacelike
$x\uu M$ directions.  The spectrum of the resulting
theory is that of type IIB in the bulk, with the following
boundary conditions at the singularity:
\bi
\item{N-type boundary conditions for
\bbb
G\ll{MN}, B\ll{MN},
G\ll{ij}, B\ll{ij}, \Phi, C\ll{iM}, C\ll{iMNP},
C\ll{ijkM},  P\ll + \up 6\Psi\uu M, P\ll -
 \up 6\Pst\uu M, P\ll - \up 6 \Psi\uu i,
{~~~\rm and ~~~}
P\ll +\up 6\Pst\uu i
\eee}
\item{D-type boundary conditions for 
\bbb
G\ll{Mi}, B\ll{Mi}, C, C\ll {MN}, C\ll{ij}, C\ll{MNPQ},
C\ll{MNij}, C\ll{ijkl},
P\ll - \up 6\Psi\uu M, P\ll + \up 6\Pst\uu M,
P\ll + \up 6 \Psi\uu i,
 {\rm ~~~ and ~~~}
P\ll -\up 6\Pst\uu i\eee}
\ei

The boundary conditions for the gravitini
indicate that the global SUSY of the 6D theory on
the singularity is of type $(1,1)$.  The localized
states do indeed organize themselves with respect to
this supersymmetry; the twisted sector contains a RR vector,
and one Weyl fermion of each chirality.  This is the content
of a vector multiplet of $(1,1)$ supersymmetry in 5+1 dimensions.

\heading{S-duality to the O5$^-$--D5 system}

In the type IIB case, we can apply S-duality to
obtain a more familiar description.  Conjugating the
$\mfls$ by S-duality, we find that the boundary condition
for bulk fields in the new description is that they are
periodic or antiperiodic according to their transformation
properties under $R\ll{6789}\cdot \Omega$, where $\Omega$
is the worldsheet parity transformation which
acts with a minus sign on $B\ll{\m\n}, C, C\ll{\m\n\s\t},$
and $\Psi - \Pst$
leaving $C\ll{\m\n}$ and all other massless NS-NS fields
invariant, including $\Psi + \Pst$.

The reader will recognize this boundary condition as the
boundary condition at an orientifold 5-plane
of the $\ttb$ theory.  However the O5 plane
cannot
be the only object present.  The amount of spactime SUSY of the 
oriented IIB theory on the $\mfls$ orbifold forbids any
corrections to the energy of the background, even
nonperturbatively.  The O5 plane 
has a nonzero tension, positive or negative
depending whether it is an O5$^+$ or an $\ofm$,
while the S-dual of
our orbifold fixed plane has zero tension.
This suggests that the dual $\ttb$ theory
must have another object to cancel the tension
of the O5 plane, namely a single D5, with the 
sign of the O5 plane constrained to be $\ofm$.  Adding a D5 to
the $\ofm$ in the $\ttb$ background also gives a
moduli space which for
small separations matches that of the infinitestimal
twisted deformations
of the IIB theory on the $\mfls$ orbifold.  That is to
say, the moduli space is four dimensional, and the deformations
transform in the ${\bf 4}$ of the transverse rotation
group.  In the $\ttb$ theory these deformations are
just the motions of the D5-brane away from the $\ofm$.

For purposes of the next section it will be interesting
to take note of the spectrum of open strings stretching
from the D5 to itself, at a point in moduli
space where the fivebrane is pulled
away from the origin by a distance
$r << \sqrt{\apr}$.  Since there is only a
single D5, the lowest open string modes stretching
from the D5, through the $\ofm$ plane to the image D5, are
projected out by the $\O$ transformation.  That is
to say, the modes which are projected out are the
ones with masses of order $r\uu 1$ for small $r$, which would
become massless with the fivebrane at the origin.  Let
us review the reason for that.

As explained in \cite{Gimon:1996rq}, the
orientifold projection
of an Op$\uu -$-plane on the oscillator ground state of the
Dp-Dp open string sectors yields states which are
antisymmetric in Chan-Paton indices.
For $k$ 
D5-branes the number of stretched open strings in
the ground state which pass through the O-plane is
$k(k-1)$, and they live in the antisymmetric
tensor representation of $U(k)$.
These ground state open strings are BPS
and live in massive vector multiplets; when the D5-brane
is coincident with the $\ofm$-plane they enhance the gauge
symmetry from $U(k)$ to $SO(2k)$.  But for $k=1$ the number of
such light open strings is zero, and going from $U(1)$
to $SO(2)$ does not enlarge
the dimension of the continuous group.

The absence of
BPS open strings stretching 
from the D5 to itself through the $\ofm$ plane
supports the S-duality between the IIB and
$\ttb$ backgrounds.  The
$U(1)$ propagating on the D5-brane of the $\ttb$ theory becomes
the $U(1)$ of the twisted Ramond-Ramond vector field 
of the IIB side.  If the gauge group were enhanced to
something nonabelian at the origin, the
continuous gauge symmetries on the two sides would not agree.

However there are still \it non-\rm BPS open strings
stretching from the D5 to itself through the $\ofm$.
An open string state can survive
the $\O$-projection if the
oscillators contribute a $-$ sign in place of the Chan-Paton
indices.  Such modes live in the rank-2 symmetric tensor
representation of the $U(k)$ group, and for a single D5 there
is one of them for each oscillator state
with $\O$ eigenvalue $-1$.  Though these strings
lack supersymmetry, the lightest is always absolutely stable.
All the stretched open string states have the same,
nonzero charge $\pm 1$ under the $U(1)$ gauge symmetry.

For small separations from the origin, the
charged stretched states do not become massless; they retain
masses $m\sqd = o({1\over{\apr}})$ due to their
oscillator energies.  
For large separations of the fivebrane
from the origin $r >> \sqrt{\apr}$
all open string states, in particular the lightest,
have masses which scale like $m\sqd \sim {{r\sqd}\over{4\pi\sqd
\apr}}$.  This formula is true at tree level, and the
locally BPS nature of the stretched strings guarantee that the
leading piece of the mass is uncorrected at strong coupling.

The presence of charged, non-BPS fundamental strings fixed
to the D5-brane in the $\ttb$ side
indicates that there should be charged, non-BPS D-branes 
pinned to the fixed locus in the IIB orbifold, the
lightest one being absolutely stable.  The presence
of a massless RR field suggests as much; in the next section
we will construct the stable non-BPS branes explicitly.

\subsection{Branes and open strings in the 
$R\ll{6789}\times (-1)\uu{F_{L_S}}$ orbifold}

\heading{General comments}

The $R\ll{6789}\times (-1)\uu{F_{L_S}}$ orbifold will
prove to be
quite similar to the Wilson line in one important
respect: the ordinary branes ('regular' branes 
in the language of \cite{Douglas:1996sw}), localized
in the transverse space away from the fixed point,
cannot be absolutely stable.  This can be seen at
many levels.  First of all, the timelike component
$C\ll 0$ of the RR vector has D-type boundary conditions
at the fixed point, meaning that it has no
zero mode in six dimensions, even one normalized
to the volume of the transverse space.  Since the
zero mode of a gauge field is the Lagrange multiplier which
enforces charge conservation, its absence suggests that
zerobrane charge need not be conserved.  Second of all,
parallel transporting a brane around the fixed point brings
it back as its own antibrane, which violates integer-valued
charge conservation.  Finally we shall see that
the zerobrane becomes unstable perturbatively
when moved sufficiently close to the fixed point.
Much of the discussion is parallel to the case of
the Wilson line for $(-1)\uu{F_{L_S}}$ and we will move through
it as tersely as possible.

On the other hand, we will see that in
the sector of branes pinned to the origin --
described in the context of the
pure $R\ll{6789}$ orbifold as 'fractional' branes -- 
there are Dp-branes with odd $p$ which
are non-BPS but \it stable \rm , unlike the usual
wrong-dimension branes of type IIA string theory.

\heading{The regular brane}

The regular brane is straightforward; we can construct it in
the usual way \cite{Douglas:1996sw},
by taking a brane and its image on
the covering space.  The inclusion of the action $\mfls$
means that the configuration on the covering space is
a brane, along with its antibrane at the
reflected point in the $X\uu i$ space.
For definiteness, consider a
D0-brane, though branes extended in some of the spacelike $X\uu M$
directions can be obtained by taking T-duals along $X\uu M$.

On a single brane, the massless bosons are a gauge field
$A\ll 0\uu {D0}$, five transverse scalars $X\uu M\ll{D0}, M\neq 0$
and four more transverse scalars $X\uu i\ll{D0}$.  There
are also sixteen fermions which transform
as a Majorana spinor of SO(9).  On an antibrane
the content is the same.  Stretching between them
is a complex scalar string $T$ whose mass is
\bbb
m\sqd \ll T =
{1\over{4\pi\sqd\apr\sqd}}
\lsqq
(X\uu M\ll{D0} - X\uu M\ll{\bar{D0}})\sqd +
(X\uu i\ll{D0} - X\uu i\ll{\bar{D0}})\sqd
\rsqq - {1\over{\apr}}
\eee
The effect of the orbifold projection is to set
$X\ll{D0}\uu M = X\ll{\bar{D0}}\uu M \equiv X\uu M$
and $X\ll{D0}\uu i = -X\ll{\bar{D0}}\uu i \equiv X\uu i$,
as well as $A\ll 0\uu{D0} = A\ll 0\uu{\bar{D0}}$.  The
tachyon $T$ transforms to its conjugate $T\st$ under
$\mfls$, so the orbifold
projection constrains the tachyon to
be real, $T = T\st$.  One linear combination
of the massless fermions is also projected out by
the orbifolding. 
This leaves a $U(1)$ gauge theory with nine
neutral scalars $X\uu M, X\uu i$, a single massless
SO(9) Majorana spinor of fermions, and a charged scalar
with mass
\bbb
m\sqd\ll T = {1\over{\pi\sqd \apr\sqd}} 
\lsqq X\uu i{}\sqd  - \pi\sqd \apr \rsqq
\eee

When the zerobrane comes within a distance $\pi\sqrt{\apr}$
of the origin, then, it becomes tachyonic and can decay. 
This is consistent with our observation that
there is no zero mode of the Ramond-Ramond
Coulomb potential $C\ll 0$.  It would be interesting
to understand the end product of the decay, whether
it is the closed string vacuum or whether some
remnant, possibly carrying a discrete charge, may
survive.  

%

\heading{Stable non-BPS branes pinned to the fixed plane}

Our theory contains a RR tensor localized to the
fixed plane $X\uu i = 0$; this suggests that the theory
should contain charged $p$-branes of odd $p$ which
are pinned to the fixed plane.  Such branes
should show up as 'fractional' branes 
in the sense of \cite{Douglas:1996sw}.
We would like to
avoid that terminology in this example, however, since
the brane which is pinned to the fixed locus
does not carry charge under any bulk RR field.

We now choose
to construct the open string Hilbert space
of the pinned D5-brane, since it preserves
the full Lorentz invariance of the
orbifold background; the D3 and D1 branes
can be obtained by T-dualities along the $X\uu M$
directions.
The boundary condition at the endpoints is
$\pp\ll n X\uu M = X\uu i = 0$ for the
bosons.  The boundary condition for the supercurrents
is $G =  \tilde{G}$ at the left endpoint, with
$G =  \pm ~ \tilde{G}$ at the right endpoint
in the R and NS sectors, respectively.
So the boundary condition on the fermions
is $\psi\uu M = \pst\uu M, \psi\uu i = -  ~\pst\uu i$
at the left endpoint, and $\psi\uu M = \pm  ~\pst\uu M,
\psi\uu i = \mp ~\pst\uu i$ at the right endpoint 
in the R and NS sectors, respectively.  

Now we would like to understand
which GSO and orbifold projections
must be imposed.  In particular, we expect that
there should not be a tachyon, since
the brane is charged and there is no obvious lighter
state which carries the same charge.

Let us take a moment to recall why stable
fivebranes are 
not allowed
in the ordinary type IIA theory, fractional or otherwise.
Starting with an open string vertex operator in
the NS$_+$ sector, we
could take its OPE with a RR vertex
operator in the bulk.  Since the bulk RR
operators have GSO projection $\rmm/\rpp$ in
type IIA, this would give an open string
with NS boundary conditions, but
GSO projection --.  Thus
one is compelled to include both values
+ and -- of $(-1)^{F_W}$ in both the NS and
R open string sectors.  In particular the open string $\nsm$ sector
contains a real tachyon.
Thus one
is left with the theory of an unstable, neutral brane
rather than
a charged, stable brane.

For the case of the type IIA orbifold
by $R\ll{6789} ~(-1)\uu{F_{L_S}}$, the 
bulk fields in the $\rpp/\rmm$ sector 
have orbifold projection $-$ rather than $+$.
This means that if we start with an open
string in the $\nsp$ sector and take its OPE
with one of the untwisted $\rmm/\rpp$ states, we
end up with an NS sector with GSO projection -
and orbifold projection --.  The matter GSO 
of such an open string is +, so the lowest
states satisfying the -- orbifold projection
are obtained by acting on the vacuum with
$X\uu i\ll{-1}$ or with 
$(\psi \uu i - \pst\uu i)\ll{-\hh}$ and $(\psi\uu M +
\pst\uu M)\ll{-\hh}$.  To get a matter state of weight
$\hh$, one must
add momentum $k\ll M$ corresponding to a
mass squared of $+ {1\over{\apr}}$.  So there are
no tachyons in the $\nsp$ sector.

More generally, total worldsheet fermion parity $(-1)\uu{F_W}$
is equal to $R\ll{6789}$-parity for \it all \rm closed
string states, twisted and untwisted.
It follows that we get a consistent
OPE between the bulk and the boundary if we apply the
same correlated projection to all open string states
as well.  The resulting sectors are:
\bi
\item{$\nsp$, with orbifold projection +.  The GSO projection
in the matter sector is then --, so
the lowest allowed states are $(\psi\uu M + \pst\uu M)\ll{-\hh}
 \kket 0$, which give the states of a massless gauge
field $a\ll M$ living on the brane.}
\item{$\nsm$, with orbifold projection --.  The GSO
projection in the matter sector is $+$, and the lowest
states surviving the projection are massive, as discussed
above.}
\item{$\rpp$, with orbifold projection +.   The
fermions all have zero modes in this sector, 
$(\psi\uu M + \pst\uu M)\ll 0$ for the M fermions
and $(\psi\uu i - \pst\uu i)\ll 0$ for the i fermions.
The projections mean that the product of the M zero modes
is +1 and the product of the i zero modes is +1.  So we
get a spacetime fermion $\m\uu A\ll p$ which transforms as
$(1,2)$ under SO(4) and as a Weyl fermion of positive
chirality under $SO(5,1)$.}
\item{$\rmm$, with orbifold projection $-$.
The
fermions all have zero modes in this sector.
The projections mean that the product of the M zero modes
is +1 and the product of the i zero modes is -1.  So we
get a spacetime fermion $\m\uu{\dot{A}}\ll p$
which transforms as
$(2,1)$ under SO(4) and as a Weyl fermion of positive
chirality under $SO(5,1)$.}
\ei
We had a choice of two GSO projections in the
open string R sectors; our choice was dictated
by the necessity that the OPE of the gravitino
vertex operator for $P\up +\Psi\uu M$
with the open string vertex operator in the NS sector
be consistent.  That is, R sector GSO is fixed by
the reqirement that the open string R sector vertex
operators contain only $\ss\ll p$'s and not $\ss\ll\pd$'s.

In addition to the stable D5,
there is also a stable D1 and D3 in this theory
which are pinned
to the origin $X\uu i = 0$.
The tree-level dynamics of the stable D3 and D1
can be obtained by performing T-dualities along
pairs of $X\uu M$ directions. 

\heading{Tadpoles, boundary states and SUSY breaking by the brane}

Earlier we claimed in passing that the pinned branes
are not charged under any bulk RR field.  This is
intuitively obvious, since there are no
bulk RR fields with the correct
Lorentz properties to couple to the 
pinned $p$-branes, but we can demonstrate
the decoupling from bulk RR fields directly,
in terms of boundary states.

The projection
on the space of open string states is $(-1)\uu{F_W}~
R\ll{6789}$, a combination of worldsheet fermion
number mod 2, and the geometric action of the reflection.
Since it is a single projection and not two independent
projections, the worldsheet partition function can be
described by a sum over two sets of boundary conditions
rather than four.  That is, the partition function
\cite{Polchinski:1998rr} at modular parameter $t$ is
\bbb
V\ll 6~\int~{{d\uu 6 k}\over{(2\pi)\uu 6}}~
\hh ~\tr\ll{{\cal H}\uu{\perp}\ll{\rm open}} \lsqq (-1)\uu{F_W}~
\lrdd 1 + (-1)\uu{F_W}R\ll{6789} \rrdd
 ~\exp{- H t} \rsqq.
\eee  The first
factor of $(-1)\uu{F_W}$ is the usual thermal
boundary condition for worldsheet fermions (it can
also be thought of as coming from the $(-1)\uu{F_W}$
contribution of the superghosts),
and the
${{1 + (-1)\uu{F_W}R\ll{6789}}
\over 2}$ implements the correlated GSO/orbifold projection.  The
trace runs over all states of the oscillators transverse
to $k\uu M$.

So the partition function
breaks up into two sectors, one with antiperiodic
boundary conditions for fermions and periodic
boundary conditions for the $X\uu \m$, and the
other sector with periodic boundary conditions for $X\uu M,
\psi\uu M,\pst\uu M$ and antiperiodic boundary conditions
for $X\uu i,\psi\uu i, \pst\uu i$.  Re-interpreted
in the closed string channel
(as in \cite{boundarystates},\cite{Polchinski:1995mt}), where 
the space and Euclidean time coordinates are reversed
relative to the open string channel, the
boundary state corresponding to the D-brane
has two contributions.  One is from the sector 
with $X\uu i$ untwisted, with NS boundary conditions for the
supercurrents.
The other is from the sector with $X\uu i$ 
antiperiodic and Ramond boundary
conditions for both supercurrents $G,\tilde{G}$.  The
target space interpretation is that the pinned
branes at the fixed locus are sources for the bulk NS/NS fields
(such as $G\ll{\m\n}$) and the twisted RR fields (such as
$A\ll{MN}$) but not the twisted NS/NS fields
(such as ${\cal M}\uu i$) or
the bulk RR fields (such as $C\ll \m$ or $C\ll{\m\n\s})$.

The absence of a one-point function for twisted
NS/NS states can be derived from symmetry as well.
All twisted NS/NS states transform in odd-rank tensor
representations of $SO(4)$, so in particular there
is no singlet which could get a tadpole.  When
we give an expectation value to the ${\cal M}\uu i$,
twisted and untwisted fields can mix in the closed
string theory, so we should expect that the
deformed D-brane state should source all possible NS/NS
and R/R fields when $\left\langle{\cal M}\uu i\right\rangle
\neq 0$.  This agrees with our picture of the
stable pinned zerobrane on the 
IIB side coming from the $\ttb$ S-dual; if the
brane is dual to a stretched excited string state,
its mass should go like $m\sim \sqrt{({\rm const.})
\cdot{\cal M}\uu i{ \cal M}\uu i + m\ll 0\sqd}$.  So the tadpole
${{\pp m}\over{\pp {\cal M}\uu i}}$ is nonzero at a generic
point in moduli space, but vanishes at ${\cal M}\uu i = 0$,
which is where we can calculate
the partition function.

We end the section with a brief comment on the supersymmetry
properties of the stable branes.  Unlike the
case of the Wilson line for $(-1)\uu{F_{L_S}}$,
the type IIA orbifold by $R\ll{6789}\cdot \mfls$ has massless
RR gauge fields propagating in six dimensions, indicating
that the odd dimensional p-branes in the singular locus
are charged and can be stable.  Indeed,
there is no tachyon in the spectrum of these branes.
Despite being stable, they are not BPS.
The only bosonic massless field propagating on the
branes is a $U(1)$ vector, but there are two sets of
massless Weyl fermions in the spectrum rather than one.
Thus it is clear that the stable pinned branes
break all the supersymmetry of the orbifold background.
In fact, our consistency condition
for the choice of R sectors
encoded the fact that the 
global SUSY of the background broken
by the brane is $(0,2)$, and 
the goldstone fermions should have the same
chirality as the corresponding gravitini.\footnote{This
in turn is opposite to the chirality of the corresponding
supercharges.}

\heading{Vertex operators}

The vertex operators for
open strings on branes work out straightforwardly.
For the regular brane, the open string Hilbert space
and vertex operators are just projections of those
for ordinary branes and antibranes on the
covering space.  This makes the
construction of vertiex operators elementary for the
regular branes and
we shall not review it.

For the stable pinned branes we now list the sectors
in the physical Hilbert space:
\bi
\item{In sector \bd a we have $\nsp$ states
(matter GSO $\nsm$).  The orbifold
projection is $+$.  The massless gauge field $a\ll M$
has vertex operator is $\psi\ll M = \pst\ll M$.}
\item{In sector \bd b we have $\rpp$ states
(matter GSO $-$).  The orbifold projection
is $+$.  The
massless state $\m\uu A\ll \pd$ has
vertex operator
$  \lim\ll{{\rm Re~z}\to 0} \ss\pr\ll \pd
\ss\ll A$.}
\item{In sector \bd c we have $\rmm$ states
(matter GSO $-$).  The orbifold projection
is $-$.  The vertex
operator for the
massless state $\m\uu\aald\ll p$
is $ \lim\ll{{\rm Re~z}\to 0} \ss\pr\ll \pd
\ss\ll\aald$.}
\item{In sector \bd d we have $\nsm$ states
(matter GSO $\nsp$).  The orbifold projection
is $-$.  There are no massless states in this sector.}
\ei

The open string vertex operators have
a consistent multiplication among themselves
and with closed string vertex operators, as
summarized by table (\ref{moreopenstringmult}).

\btt
\begin{center}
\bta{|c||c|c|c|c|}
\rowf {} a b c d \\
\rowf a a b c d \\
\rowf b b a d c \\
\rowf c c d a b \\
\rowf d d c b a  \\
\hline
\hline
\rowfa A H a b c d \\
\rowfa B G b a d c \\
\rowfa C F c d a b \\
\rowfa D E d c b a \\
\hline\hline
\eta
\caption{Multiplication rules for open and closed
string vertex operators in the background of the
stable pinned brane at the fixed locus of the
orbifold by $R\ll{6789}$ with an action of
$\mfls$.}
\label{moreopenstringmult}
\end{center}
\ett

\heading{Type IIB version}

The branes of the type IIB theory on the $\mfls~R\ll{6789}$
orbifold can be understood simply from T-dualizing
the type IIA version, along an odd number of the spacelike
$x\uu M$ directions.  The behavior of the IIB branes
is, not surprisingly, parallel to that of the IIA branes.  The
regular branes are perturbatively stable far from the
fixed point, developing an instability when they 
come within a distance $\pi\sqrt{\apr}$ of the origin.
There are fractional zerobranes and twobranes
which are electrically and magnetically charged
under the twisted RR vector multiplet, as well as
a fourbrane which is charged under the (nondynamical)
twisted RR five-form.
These branes are all non-BPS.  Since they
are the lightest objects charged under their respective
gauge potentials, they must be absolutely stable,
not just perturbatively.

The existence of the non-BPS
pinned zerobrane in type IIB orbifold background
is an interesting check on the duality to the $\ofm$+D5
system in the type $\ttb$ dual.  As we pointed out
earlier the D5, when separated from the $\ofm$, has
open strings stretching from itself, through the $\ofm$-plane
to itself again.  These strings never become massless,
even when the D5 returns to the origin.  They are charged
under the gauge field on the D5 and the lightest is
absolutely stable.  S-duality then dictates the existence
of a stable, non-BPS object charged under the twisted
RR one-form.  The stable non-BPS zerobrane of
IIB on the $\mfls$ orbifold has exactly the right properties
to be the dual state.

\section{Discussion}

In this paper we have presented two backgrounds of type II
string theory which are exactly solvable at tree
level and preserve 16 supercharges.  Both 
backgrounds display new phenomena
and illuminate unexpected aspects
of highly supersymmetric theories of quantum gravity.  We have
focused on type IIA, but the properties
of type IIB on the same
backgrounds can always be studied by performing
a T-duality along one of the trivial directions.

The first background, a Wilson line
for $\mfls$, displays
a self-duality -- as opposed to a duality to
the opposite type II theory -- on a single circle, as well
as enhanced
gauge symmetry at the self-dual radius.  In type 
IIA string theory, the operation $\mfls$ is simply a
reflection of the M-circle (together with the operation
$C\ll{\m\n\s} \to - C\ll{\m\n\s}$ on the M-theory three-form),
so the Wilson line background can be interpreted as M theory on
a Klein bottle.  Given that interpretation, the
self-duality and appearance of enhanced gauge symmetry are
quite surprising and new.

We have been forced to conclude that
M-theory on a Klein bottle has a self-duality which
changes the volume of the Klein bottle and maps wrapped
twobranes into Kaluza-Klein modes.  The duality is remaniscent
of the self-duality of M-theory on a $T\uu 3$
\cite{Sen:1995cf}, but logically
distinct from it, nor can the former be derived from
the latter.  Since the duality maps
ordinary, locally charged D-branes to unstable, uncharged
branes, this connection might be used to
shed light on the place of non-BPS branes in M-theory.

Our second background is type II string theory on
$\IR\uu 4 / \IZ\ll 2$, where the $\IZ\ll 2$
acts as $\mfls$ in addition to the geometric
action $R\ll{6789}$.
This background has the field content of
type IIA/B in the bulk, but the localized
modes and fractional branes
are those of type IIB/A.  In neither case
are the 'regular' branes, pointlike and separated from the
fixed point, absolutely stable, though sufficiently far
from the fixed point they are tachyon-free at tree level.

These orbifolds have twisted marginal operators
which can be integrated to finite marginal directions.
The nature of the marginal deformation of the CFT is
unknown, but is possibly nongeometric in the sense of
\cite{Hellerman:2002ax}.  Indeed, the
orbifold arises in a certain decompactification limit
of the $\hat{c} = 4$ example worked out in that
paper. 

In type IIB, the 
marginally deformed orbifold
can be S-dualized to a D-fivebrane pulled away from
an $\ofm$-plane.  Therefore the deformed orbifold
contains an
NS-fivebrane and has nonvanishing H-flux.
The direct interpretation in type IIA is far from clear,
though the NS/NS content of the IIA version
and the IIB version must be the
same.  The presence of H-flux in the deformed orbifold
agrees with what one would generally expect
from the $T\uu 2$ fibration
picture of 
\cite{Hellerman:2002ax}.  In those examples,
H-flux always appears when one deforms
away from the asymmetric orbifold point.
However the decompactification does not
commute straightforwardly with the deformation
of the orbifold, since the volume
of the $T\uu 2$ fiber is not a free parameter in the
compact deformed orbifold.  Further study
would help to make the connection with nongeometric
string theories more completely clear.

\vskip.5in

\centerline{\bf{Acknowledgements}}

The author would like to thank the
Department of Energy for support under
grant DE-FG02-90ER40542.  He would also
like to thank Ra\'ul Rabadan for 
comments on the draft, and especially
Johannes Walcher for
valuable discussions and collaboration on
related work.  The author is the D. E. Shaw \& Co., L. P. Member
at the Institute for Advanced Study.


\begin{thebibliography}{99}



\small


\bibitem{Hellerman:2004zm}
  S.~Hellerman,
  ``On the landscape of superstring theory in D $>$ 10,''
  {\tt hep-th/0405041}.

\bibitem{Polchinski:1998rr}
  J.~Polchinski,
  ``String theory. Vol. 2: Superstring theory and beyond,''

\bibitem{Horava:1998jy}
  P.~Horava,
  ``Type IIA D-branes, K-theory, and matrix theory,''
  Adv.\ Theor.\ Math.\ Phys.\  {\bf 2}, 1373 (1999)
  {\tt hep-th/9812135}.


\bibitem{Gimon:1996rq}
  E.~G.~Gimon and J.~Polchinski,
  ``Consistency Conditions for Orientifolds and D-Manifolds,''
  Phys.\ Rev.\ D {\bf 54}, 1667 (1996)
  {\tt hep-th/9601038}.

\bibitem{Green:1987sp}
  M.~B.~Green, J.~H.~Schwarz and E.~Witten,
  ``Superstring Theory. Vol. 1: Introduction.''

\bibitem{Hellerman:2002ax}
  S.~Hellerman, J.~McGreevy and B.~Williams,
  ``Geometric constructions of nongeometric string theories,''
  JHEP {\bf 0401}, 024 (2004)
  {\tt hep-th/0208174}.

\bibitem{Witten:1995ex}
  E.~Witten,
  ``String theory dynamics in various dimensions,''
  Nucl.\ Phys.\ B {\bf 443}, 85 (1995)
  {\tt hep-th/9503124}.

\bibitem{Douglas:1996sw}
  M.~R.~Douglas and G.~W.~Moore,
  ``D-branes, Quivers, and ALE Instantons,''
  {\tt hep-th/9603167}.

\bibitem{related}
  A.~Dabholkar and C.~Hull,
  ``Generalised T-Duality and Non-Geometric Backgrounds,''
  {\tt hep-th/0512005} ; \\
  A.~Flournoy and B.~Williams,
  ``Nongeometry, duality twists, and the worldsheet,''
  {\tt hep-th/0511126}; \\
  J.~Shelton, W.~Taylor and B.~Wecht,
  ``Nongeometric flux compactifications,''
  JHEP {\bf 0510}, 085 (2005)
  {\tt hep-th/0508133}; \\
  J.~Gray and E.~Hackett-Jones,
  ``On T-folds, G-structures and supersymmetry,''
  {\tt hep-th/0506092}; \\
  C.~M.~Hull and R.~A.~Reid-Edwards,
  ``Flux compactifications of string theory on twisted tori,''
  {\tt hep-th/0503114}; \\
  C.~M.~Hull,
  ``A geometry for non-geometric string backgrounds,''
  JHEP {\bf 0510}, 065 (2005)
  {\tt hep-th/0406102}; \\
  A.~Flournoy, B.~Wecht and B.~Williams,
  ``Constructing nongeometric vacua in string theory,''
  Nucl.\ Phys.\ B {\bf 706}, 127 (2005)
  {\tt hep-th/0404217}; \\
  A.~Dabholkar and C.~Hull,
  ``Duality twists, orbifolds, and fluxes,''
  JHEP {\bf 0309}, 054 (2003)
  {\tt hep-th/0210209}.

\bibitem{Friedan:1985ge}
  D.~Friedan, E.~J.~Martinec and S.~H.~Shenker,
  ``Conformal Invariance, Supersymmetry And String Theory,''
  Nucl.\ Phys.\ B {\bf 271}, 93 (1986).

\bibitem{ssc}
  R.~Rohm,
  ``Spontaneous Supersymmetry Breaking In Supersymmetric String Theories,''
  Nucl.\ Phys.\ B {\bf 237}, 553 (1984); \\
  N.~Seiberg,
  ``Observations on the moduli space of two dimensional string theory,''
  JHEP {\bf 0503}, 010 (2005)
  {\tt hep-th/0502156};\\
  C.~Angelantonj and A.~Sagnotti,
  ``Open strings,''
  Phys.\ Rept.\  {\bf 371}, 1 (2002)
  [Erratum-ibid.\  {\bf 376}, 339 (2003)]
  {\tt hep-th/0204089}.

\bibitem{boundarystates}
C. Lovelace, Phys. Lett. {\bf B34}, 500 (1971);\\
L. Clavelli and J. Shapiro, Nucl. Phys. {\bf B57}, 490 (1973);\\
M. Ademollo, R. D' Auria, F. Gliozzi, E. Napolitano, S. Sciuto, and P. di
Vecchia, Nucl. Phys. {\bf B94}, 221 (1975);\\
C. G. Callan, C. Lovelace, C. R. Nappi, and S. A. Yost,
Nucl. Phys. {\bf B293}, 83 (1987); \\
J. Polchinski and Y. Cai, Nucl. Phys. {\bf B296}, 91 (1988);\\
C. G. Callan, C. Lovelace, C. R. Nappi and S.A. Yost,
Nucl. Phys.  {\bf B308}, 221 (1988); \\
M. Bianchi and A. Sagnotti, Phys. Lett. {\bf 247B}
(1990) 517; Nucl. Phys. {\bf B361} (1991) 519;\\
P. Horava, Nucl. Phys. {\bf B327}, 461 (1989).


\bibitem{Polchinski:1995mt}
  J.~Polchinski,
  `` Dirichlet-Branes and Ramond-Ramond Charges'',
  Phys.\ Rev.\ Lett.\  {\bf 75}, 4724 (1995)
  {\tt hep-th/9510017}.

\bibitem{Sen:1995cf}
  A.~Sen,
  ``T-Duality of p-Branes,''
  Mod.\ Phys.\ Lett.\ A {\bf 11}, 827 (1996)
  {\tt hep-th/9512203}.

\bibitem{sens}
A.~Sen,
  ``Duality and Orbifolds,''
  Nucl.\ Phys.\ B {\bf 474}, 361 (1996)
  {\tt hep-th/9604070}.



\bibitem{oren}
O.~Bergman and M.~R.~Gaberdiel,
  ``Stable non-BPS D-particles,''
  Phys.\ Lett.\ B {\bf 441}, 133 (1998)
  {\tt hep-th/9806155}.

\bibitem{trunc}
  M.~R.~Gaberdiel and B.~J.~Stefanski,
  ``Dirichlet branes on orbifolds,''
  Nucl.\ Phys.\ B {\bf 578}, 58 (2000)
  {\tt hep-th/9910109}.

\end{thebibliography}
\end{document}